\documentclass{IEEEtran}
\usepackage{xcolor,soul,framed} 
\usepackage[noadjust]{cite}
 
\colorlet{shadecolor}{yellow}
\usepackage[pdftex]{graphicx}
\graphicspath{{../pdf/}{../jpeg/}}
\DeclareGraphicsExtensions{.pdf,.jpeg,.png}

\usepackage{amssymb,mathtools}
\usepackage{array}
\usepackage{mdwmath}
\usepackage{mdwtab}
\usepackage{eqparbox}
\usepackage{url}
\usepackage[ruled,vlined,linesnumbered]{algorithm2e}
\usepackage{enumitem}
\usepackage[mathcal]{euscript} 
\usepackage{tabularx}
\usepackage{multirow} 
\usepackage{booktabs} 
\usepackage{makecell} 
\usepackage{aligned-overset} 
\usepackage{url}
\usepackage{multicol} 

\usepackage{balance}
\usepackage{bbm}
\usepackage{amsmath}
\usepackage{amsthm}

\usepackage{pifont}
\newcommand{\cmark}{\textcolor{green!80!black}{\ding{52}}}
\newcommand{\xmark}{\textcolor{red}{\ding{55}}}

\graphicspath{{./Figures/}}
\newtheorem{Remark}{Remark}
\newtheorem{Lemma}{Lemma}
\newtheorem{Definition}{Definition}
\newtheorem{Theorem}{Theorem}

\renewcommand{\vec}[1]{\boldsymbol{\mathrm{#1}}}

\newcommand{\s}{\mathrm{T}}

\newcommand{\tx}{\mathrm{T}}
\newcommand{\rx}{\mathrm{R}}
\newcommand{\IRS}{\mathrm{I}}
\newcommand{\ue}{\mathrm{R}}

\newcommand{\SINR}{\mathrm{SINR}}

\newcommand{\icd}{\mathrm{inc}}
\newcommand{\rfl}{\mathrm{rfl}}

\newcommand{\pl}{\xi}
\newcommand{\gain}{\mathbb{G}}

\usepackage[left=.5in,                               right=.5in,                     top=0.9in,  bottom=0.9in]{geometry}

\begin{document}

    \title{{Multi-IRS-aided Terahertz Networks: Channel Modelling and User Association With Imperfect CSI}}
   \author{
  {Muddasir Rahim\IEEEauthorrefmark{1}, Thanh Luan Nguyen\IEEEauthorrefmark{1}, Georges~Kaddoum\IEEEauthorrefmark{7},~\IEEEmembership{Senior~Member,~IEEE}, and Tri~Nhu~Do\IEEEauthorrefmark{2}}
  \thanks{\IEEEauthorrefmark{1}M.~Rahim and T.~L.~Nguyen are with the Department of Electrical Engineering, \'{E}cole de Technologie Sup\'{e}rieure (\'{E}TS), Universit\'{e} du Qu\'{e}bec, Montr\'{e}al, QC H3C 1K3, Canada (emails: muddasir.rahim.1@ens.etsmtl.ca,~thanh-luan.nguyen.1@ens.etsmtl.ca,).

 \IEEEauthorrefmark{7}G.~Kaddoum is with Department of Electrial Engineering, \'{E}cole de Technologie Sup\'{e}rieure (\'{E}TS), Universit\'{e} du Qu\'{e}bec, Montr\'{e}al, QC H3C 1K3, Canada and Artificial Intelligence \& Cyber Systems Research Center, Lebanese American University (email: georges.kaddoum@etsmtl.ca)
  
  \IEEEauthorrefmark{2}T.~N.~Do is with the Department of Electrical Engineering, Polytechnique Montreal, QC H3T 1J4, Canada (email: tri-nhu.do@polymtl.ca).} }
\maketitle
\begin{abstract}
  Terahertz (THz) communication is envisioned as one of the candidate technologies for future wireless communications to enable achievable data rates of up to several terabits per second (Tbps). However, the high pathloss and molecular absorption in THz band communications often limit the transmission range. To overcome these limitations, this paper proposes intelligent reconfigurable surface (IRS)-aided THz networks with imperfect channel state information (CSI). Specifically, we present an angle-based trigonometric channel model to facilitate the performance evaluation of IRS-aided THz networks. In addition, to maximize the sum rate, we formulate the transmitter (Tx)-IRS-receiver (Rx) matching problem, which is a mixed-integer nonlinear programming (MINLP) problem. To address this non-deterministic polynomial-time hard (NP-hard) problem, we propose a Gale-Shapley algorithm-based solutions to obtain stable matching between transmitters and IRSs, and receivers and IRSs, in the first and second sub-problems, respectively. The impact of the transmission power, the number of IRS elements, and the network area on the sum rate are investigated. Furthermore, the proposed algorithm is compared to an exhaustive search, nearest association, greedy search, and random allocation to validate the proposed solution. The complexity and convergence analysis demonstrate that the computational complexity of our algorithm is lower than that of the ES method.
\end{abstract}

\section{INTRODUCTION}\label{introduction}
\IEEEPARstart{T}{he} exponential growth in the number of connected devices and multimedia applications have dramatically increased the demand for large bandwidth (BW) and high data rate transmissions~\cite{rappaport2019wireless}. To accommodate these demands, the possible use of the terahertz (THz) band has attracted great interest from both industry and academia. The THz band can provide large available BW (from $0.1$ THz to $10$ THz) and a higher data rate of up to 1 terabits-per-second (Tbps), compared with the millimeter-wave (mmWave) band~\cite{chowdhury20206g,ieee2017ieee}. In our model, we aim to improve the network's sum rate. In this context, the THz band is considered to be a promising candidate to enable beyond fifth-generation (B5G) and sixth-generation (6G) wireless communications~\cite{wan2021terahertz}.

The THz band has many advantages, but establishing a reliable transmission link at THz frequencies is not easy~\cite{han2016distance}. The reason is that there are strong atmospheric attenuations, extremely high free-space losses, and the line-of-sight (LOS) channel is highly sensitive to blockage effects. This may negatively affect the communication range and reduce service coverage of THz communication networks. In the context of THz communications, deploying ultra-massive multiple-input-multiple-output (MIMO) systems may provide significant signal gains and potentially overcome the mentioned problem, however, ultra-massive MIMO requires a huge power source~\cite{sarieddeen2019terahertz}.

Recently, intelligent reconfigurable surfaces (IRSs) have become a promising technology to tackle the above problem of THz networks \cite{DoTCOMM2021}. IRSs are made of a massive number of small passive and metamaterial-based reconfigurable elements, and an IRS can manipulate both the phase and amplitude of incident signals to reflect the signals in the desired directions. In addition, IRSs can be utilized to improve the performance of wireless communication systems, including for coverage extension, confidentiality improvement, and fairness guarantee. Specifically, the use of IRSs is useful when the LOS channel is blocked or has weak received signal power since it is possible to provide additional transmission links by utilizing the reflecting elements of an IRS. 

IRS can be considered as passive IRS or active IRS. In the case of passive IRS, no power is needed when it manipulates the incident signal with the fixed reflecting coefficients. However, some power-aided circuits are deployed at the IRS to adjust the PS of each IRS element, which can be tens of microwatts per reflecting element~\cite{huang2019reconfigurable}. In the case of active IRS, each reflecting element is supported by an active load impedance. The active IRS is similar to an active reflector, which adjusts the PS as well as performs the power amplification, with a less complex and power-hunger radio frequency (RF) chain~\cite{long2021active}. However, signal amplification at the active IRS also amplifies the received noise at the IRS, ultimately increasing the noise power at the receiver. Furthermore, the active IRS is suitable for space-limited cases where deploying massive reflecting elements is difficult~\cite{long2021active}. Moreover, unlike other transmission technologies such as active relay~\cite{di2020reconfigurable} IRS-aided networks are cost-efficient and energy-efficient because the IRS does not need power amplifiers and is composed of passive reflecting elements. Based on the benefits of THz bands and IRSs, we introduce three potential application frameworks, namely, IRS-aided outdoor communications (i.e., dense building outdoor scenarios), IRS-aided indoor communications, and IRS-aided unmanned aerial vehicle (UAV) communications~\cite{diamanti2021prospect}. Specifically, the IEEE 802.15.3d standard selected the frequency range of 0.252-0.325 THz to support applications that require tens of gigabits per second (Gbps) in data rate~\cite{ieee2017ieee}.

Compared to single IRS-aided wireless networks, multiple IRS-aided wireless networks were shown to be more robust and capable of providing a greater range of services~\cite{yang2021energy}. However, IRSs may have limited computing capabilities to support signal processing, especially when heterogeneous IRS deployment in a given network area is considered. In such a scenario, user pairing is essential for maximizing the IRS's efficiency. In addition, to maximize the sum rate through multiple IRS-aided wireless networks, a fast and efficient IRS scheduling policy is vital. Due to the binary association variables, the IRS association problems are generally non-deterministic polynomial-time hard (NP-hard) in general, which necessitates proper algorithm design. Meanwhile, there are two potential approaches. One is to relax the binary variables to continuous ones and use the Lagrangian dual decomposition scheme, as in~\cite{liu2019user}. However, the relaxed variables may result in loose upper-bound solutions that are not sufficiently accurate for the original problems. The second approach is to treat the association problem as a matching game. Motivated by these facts, we propose a low-complexity matching-based IRS scheduling scheme, i.e., the scheduling from each transmitter to an IRS and from that particular IRS to a receiver, that maximizes the sum rate of the entire network.

\subsection{Related Works}\label{relatedworks}
Existing works on IRS-aided networks mostly focus on the phase shift (PS) matrix with passive beamforming at the IRS to maximize the signal-to-noise ratio (SNR), sum rate, secrecy rate, or energy efficiency~\cite{he2020coordinated, xiu2020secure, xie2022robust,zhang2021beyond}, without considering IRS selection. A distributed IRS-aided wireless network was proposed in~\cite{he2020coordinated} to maximize achievable sum rates by optimizing the PS matrix with passive beamforming at all distributed IRSs and the transmit power vector at the sources. Moreover, the use of multiple IRS-aided mmWave communications to maximize the secrecy rate has been investigated in~\cite{xiu2020secure} and proposed a joint optimization of IRS PSs, transmit beamforming, and IRS on-off status. In~\cite{xie2022robust}, the authors investigated the robust joint design of the IRS-assisted cell-free MIMO communication network, where the objective was to maximize the average sum rate by jointly designing the active transmit beamforming of the APs and passive reflecting beamforming of the IRS. The authors in~\cite{zhang2021beyond} demonstrated the energy efficiency of IRS-aided cell-free MIMO networks. In this context, the authors proposed a hybrid beamforming technique consisting of digital beamforming at APs and analog beamforming on the IRS to maximize energy efficiency. The authors in~\cite{hou2020reconfigurable} proposed a prioritized signal-enhancement-based (SEB) method, where passive beamforming is designed for the user with the best channel gain, and all the other users rely on IRS-enhanced beamforming.

There are few studies that addressed the problem of IRS-user scheduling in multiple-input-single-output (MISO) networks.
The authors in~\cite{fang2020optimum} proposed IRS selection strategies for multi-IRS-aided wireless networks in which the number of elements of each IRS can be arbitrarily set. The performance analysis was carried out based on assuming that the magnitudes of the channel coefficients associated with different IRSs are independent and identically distributed (IID) random variables. In~\cite{mei2020cooperative}, the authors proposed an IRS selection policy that maximizes the end-to-end (e2e) SNR for multi-IRS-aided networks. In this context, the authors only consider the impact of pathloss (PL) and ignore channel fading effects. The effect of distributed and centralized IRS deployment schemes on the capacity region of multi-IRS-aided wireless networks has been demonstrated in~\cite{zhang2020intelligent}. According to the authors, in distributed IRS deployments, the channels associated with each distributed IRS are subject to IID Rayleigh fading.

The authors in~\cite{yildirim2020modeling} considered multi-IRS-aided wireless networks for both outdoor and indoor communications, where the direct channel between a transmitter and a receiver is blocked. A low-complexity IRS selection scheme that uses the IRS with the highest SNR for communication is proposed in this study. However, small-scale fading was not considered, and no performance analysis was performed. In~\cite{jung2021optimality}, the authors derived the asymptotic optimal solution for passive beamforming to maximize the achievable sum rate of the IRS-assisted network. Furthermore, this work proposed a joint user association and transmit power control in a multi-user downlink network to maximize the sum rate of the network. The authors in~\cite{mei2021performance} formulated the optimization problem to maximize the minimum average signal-to-interference-plus-noise ratio (SINR) among all users by optimizing the IRS-user allocations. The nearest association (NA) method is proposed, where each user is associated with the IRS which is closest to it among all IRSs. In~\cite{sun2023irs}, the authors considered an IRS-aided cellular-based internet-of-things (IoT) system model where the IoT devices are powered by the ambient radio frequency (RF) energy from cellular transmissions. Moreover, the NA method is used for IoT devices to the APs association and for IoT devices to the IRSs association.
\begin{table}[!t]
        \centering
        \scriptsize
       {\caption{Comparison of Related Work on IRS-Assisted Communications }}
        \begin{tabular}{|c|c|c|c|c|c|c|c|}     \hline                            Ref.& DN & MI & IA & PSC & THz & M & Performance metrics\\ \hline\hline        \cite{he2020coordinated}  & \cmark & \cmark & \xmark &\cmark&\xmark & Small & Sum rate \\ \hline               \cite{xiu2020secure} &\cmark & \cmark &\xmark &\cmark & \xmark & Small & Secrecy rate  \\ \hline            \cite{xie2022robust} &\cmark & \cmark &\xmark &\cmark & \xmark & Small & Sum rate   \\ \hline           \cite{zhang2021beyond} &\cmark & \cmark &\xmark &\cmark & \xmark & Small & Energy efficiency \\ \hline           {\cite{hou2020reconfigurable}} &\cmark & \cmark &\xmark &\cmark & \xmark & Small &Outage probability    \\ \hline           \cite{fang2020optimum} &\cmark & \cmark &\cmark &\xmark & \xmark & Small &Outage probability  \\ \hline           \cite{mei2020cooperative} &\cmark & \cmark &\cmark &\xmark & \xmark & Large & Channel gain \\ \hline          \cite{yildirim2020modeling} &\cmark & \cmark &\cmark &\xmark & \xmark & Large &Bit error rate \\ \hline
          {\cite{jung2021optimality}} &\cmark & \cmark &\cmark &\cmark & \xmark & Large & Sum rate  \\ \hline
         {\cite{mei2021performance}} &\cmark & \cmark &\cmark &\xmark & \xmark &  Large & {Average SINR}   \\ \hline 
         {\cite{sun2023irs}} &\cmark & \cmark &\cmark &\xmark & \xmark &  Small & {Average throughput}   \\ \hline
          Proposed&\cmark & \cmark &\cmark &\cmark & \cmark & Very large & Sum rate  \\ \hline 
         \end{tabular} 
          
          {\vspace{5pt} \begin{tabular}[]{@{}l@{}} Distributed network (DN), multiple IRS (MI), IRS association (IA),\\PS configuration (PSC), Number of reflecting elements (M).\end{tabular}}         \label{tab:comp}           
    \end{table}
\subsection{Contributions}\label{contributions}
In contrast to the aforementioned related works, this paper considers the Tx-IRS-Rx scheduling problem in IRS-aided networks as a matching problem. To the best of our knowledge, this is the first paper to study the Tx-IRS-Rx allocation problem in IRS-aided THz networks in the presence of imperfect channel state information (CSI) knowledge in order to improve the sum rate. A comparison of existing works on IRS-assisted communications and the novelty of the proposed work is summarized in Table~\ref{tab:comp}. The key contributions of this paper are summarized as follows: 

\begin{itemize}
  \item In this paper, we derive the channel model for IRS-aided THz networks without direct channels. We also propose a new angle-based trigonometric PL model and compare it to existing models to validate its correctness.
     \item In order to maximize the sum rate, we propose a method to formulate the allocation problem as a three-dimensional (3D) matching problem. Furthermore, we mathematically prove that the formulated problem is NP-hard. We then propose a global optimal solution for the formulated 3D allocation problem by leveraging the exhaustive search (ES) method as a benchmark.
  \item For the formulated association problem, we first decompose the 3D problem into two 2D sub-problems. We then propose two consecutive Gale-Shapley algorithms for the decomposed problems to obtain a sub-optimal solution to the original problem. Moreover, the proposed sub-optimal solution is near-optimal and significantly reduces the computational complexity. 
    \item We prove that the proposed matching-based algorithm converges to a stable matching and terminates after a finite number of iterations.
    \item Numerical results demonstrate that the proposed matching-based allocation can significantly improve the sum rate compared to the NA, greedy search (GS), and random assignment (RA). Additionally, the sum rate of the proposed method is compatible with the sum rate of the ES method, while the computational complexity of the proposed method is lower than the ES method. 
\end{itemize}
\subsection{Paper Organization and Notation}
The remainder of the paper is outlined as follows. First, the system model, including the IRS PS configurations, channel modeling, and the SINR formulation, is illustrated in Section~\ref{model}. The IRS association optimization problem is formulated in Section~\ref{pform}. In Section~\ref{optimal_solution}, an optimal solution is provided, and the proposed low-complexity algorithm is presented in Section~\ref{proposed}. The performance of the proposed approach is examined in Section~\ref{pref}, and the paper is concluded in Section~\ref{conc}.

{\it Notations: } Boldface lowercase and uppercase symbols are used for vectors and matrices, respectively. The symbols $\vec{X}^{\sf T}$ and $\vec{X}^{\sf H}$ denote the transpose and conjugate transpose (Hermitian) of matrix $\vec{X}$, respectively. We use $[\vec{x}]_n$ and $[\vec{X}]_{n,m}$ to represent the $n$-th and $(n,m)$-th elements of the respective vector $\vec{x}$ and matrix $\vec{X}$, respectively. The notations $|.|$ and $\Vert . \Vert$ are the absolute value and L2-norm operators, respectively. ${\mathbb{E}}[.]$
denotes the statistical expectation. $\mathbb{R}^{M \times 1}$ and $\mathbb{C}^{M \times 1}$ denote the set of real and complex vectors of length $M\times 1$, respectively. We use $\mathcal{CN}(\mu,\sigma^2)$ to define a circularly symmetric complex Gaussian (CSCG) random variable with mean $\mu$ and variance $\sigma^2$. Finally, $\vec{X}^\star$ and $\overline{\vec{X}}$ denote the optimal and near-optimal values of the optimization variable $\vec{X}$, respectively. Important system parameters and channel variables used in this paper are detailed in Table~\ref{tab:para}.
 \begin{figure}[!t]
 \centering
 \includegraphics[width=\linewidth]{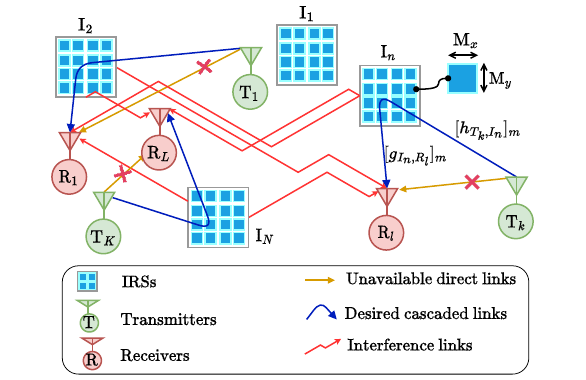}
 \caption{IRS-aided networks over THz band.
 }
 \label{m1}
 \end{figure}
 \begin{table}[t]\label{tab:para}
 \caption{{List of mathematical notations}}
  \begin{tabular}{l l }
 \toprule \textbf{Notation} & \textbf{Description}            \\ \hline
${K}$  &  Number of transmitters. \\ 
${L}$  &  Number of receivers. \\
${N}$  &  Number of IRSs. \\
${M}$  &  Number of IRS reflecting elements. \\
$\tx_k$&$k^{th}$ transmitter.\\
$\rx_l$&$l^{th}$ receiver.\\
$\IRS_n$&$n^{th}$ IRS.\\
$\IRS_{n,m}$&$m^{th}$ reflecting element on $\IRS_n$.\\
$\vec c_{\IRS}^n$& Center location of $\IRS_n$.\\
$\vec c_{\tx}^k$& Center location of $\tx_k$.\\
$\vec c_{\rx}^l$& Center location of $\rx_l$.\\
$d_{\tx_k,\IRS_n}$ &Distance between $\tx_k$ to $\IRS_n$.\\
$d_{\IRS_n,\rx_l}$ &Distance between $\IRS_n$ to $\rx_l$.\\
$\psi$& Evaluation angle.\\
$\phi$& Azimuth angle.\\
$x_k$& Transmit signal of transmitter $k$.\\
${\widehat{\vec{h}}}_{\tx_k,\IRS_n}{ \in \mathbb{C}}^{M \times 1}$& EC vector from $\tx_k$-to-$\IRS_n$.\\
${\widetilde{\vec{h}}}_{\tx_k,\IRS_n}$& CEErs from $\tx_k$-to-$\IRS_n$.\\
${\vec{h}}_{\tx_k,\IRS_n}{ \in \mathbb{C}}^{M \times 1}$& Channel vector from $\tx_k$-to-$\IRS_n$.\\
${\widehat{\vec{g}}}_{\IRS_n,\rx_l}{ \in \mathbb{C}}^{M \times 1}$& EC vector from $\IRS_n$-to-$\rx_l$.\\
${\widetilde{\vec{g}}}_{\IRS_n,\rx_l}$& CEEs from $\IRS_n$-to-$\rx_l$.\\
${\vec{g}}_{\IRS_n,\rx_l}{ \in \mathbb{C}}^{M \times 1}$& Channel vector from $\IRS_n$-to-$\rx_l$.\\
$h_{\tx_k,\IRS_n,\rx_l}$& Cascaded channel from $\tx_k$-to-$\rx_l$ through $\IRS_n$.\\
$\widehat{h}_{\tx_k,\IRS_n,\rx_l}$& Cascaded EC from $\tx_k$-to-$\rx_l$ through $\IRS_n$.\\
$\widetilde{h}_{\tx_k,\IRS_n,\rx_l}$& CEE from $\tx_k$-to-$\rx_l$ through $\IRS_n$.\\
$\pl_{\tx_k,\IRS_{n,m}}$& Pathloss in $\tx_k\to\IRS_{n,m}$ link.\\
$\pl_{\IRS_{n,m},\rx_l}$& Pathloss in $\IRS_{n,m}\to\rx_{l}$ link.\\
$\pl_{\tx_k\IRS_{n,m},\rx_l}$& Pathloss in $\tx_k\to\IRS_{n,m}\to\rx_{l}$ link.\\
$\vec{\Theta}_n$& IRS reflection matrix.\\

$\kappa_{n,m} \in (0,1]$ &Reflection coefficient of $\IRS_{n,m}$.\\
$\theta_{n,m} \in [0, 2\pi)$&Phase shift applied by $\IRS_{n,m}$.\\
$p_k$& Transmit power of $\rx_k$.\\
$\vec{r}_{k,n,m}$& Incident direction from $\tx_k$-to-$\IRS_{n,m}$.\\
$\vec{r}_{n,m,l}$& Reflection direction from $\IRS_{n,m}$-to-$\rx_l$.\\
$\gain_{\IRS_{n,m}}(-\vec{r}_{k,n,m})$ &Gains of $\IRS_{n,m}$ in the incident direction $\vec{r}_{n,m,l}$.\\
$\gain_{\IRS_{n,m}}(\vec{r}_{n,m,l})$ &Gains of $\IRS_{n,m}$ in the reflection direction $\vec{r}_{n,m,l}$.\\
$\gain_{\tx_k}$ & Antenna gain at $\tx_k$.\\
$\gain_{\rx_l}$ &Antenna gain at $\rx_l$.\\
$y_{\tx_k,\IRS_{n},\rx_l}$& Received signal at $\rx_l$ from $\tx_k$ through $\IRS_n$.\\
$[\vec{\Omega}]^\star_{\tx_k,\IRS_n,\rx_l}$ &Allocation matrix obtained with ES method.\\
$[\overline{\vec{\Omega}}]_{\tx_k,\IRS_n,\rx_l}$ &Allocation matrix obtained with proposed algorithm.\\
$\SINR_{\tx_k,\IRS_n,\rx_l}$& $\SINR$ from $\tx_k\to\IRS_n\to\rx_l$ link.\\
$R_{\tx_k,\IRS_n,\rx_l}$& Data rate of $\tx_k\to\IRS_n\to\rx_l$ link.\\
$\Xi_{\tx_k,\IRS_n , :}$&Pseudo $\SINR$s from $\s_k\to \IRS_n$ link.\\
$\Xi_{{\tx_{k^\star}},{\IRS_{n^\star}},{\rx_l}}$ & Pseudo $\SINR$s from $\s_{k^\star}\to \IRS_{n^\star}\to \ue_l$ link. \\
$ \Lambda_{\tx_k,\IRS_n , :}$& Pseudo data rate of $\tx_k\to\IRS_n$ link.\\
$\Lambda_{\tx_{k^\star},\IRS_{n^\star}, \rx_l}$& Pseudo data rate of $\tx_{k^\star}\to\IRS_{n^\star}\to\rx_l$ link.\\
\bottomrule
 \end{tabular}
 \end{table}
\section{System Model} \label{model}
We consider IRS-aided networks using THz bands where the IRSs are used to assist the communication between the transmitter (Tx) and the receiver (Rx). The network consists of a set of available Txs $\mathcal{T} = \{\tx_1,\tx_2,\ldots, \tx_{{K}} \}$ and a set of distributed Rxs $\mathcal{\rx} = \{\rx_1,\rx_2,\ldots, \rx_{{L}} \}$, with a cardinality of ${K}$ and ${L}$, respectively. The locations of
the nodes are assumed to be random but uniformly
distributed within a defined region. Furthermore, we assume that the reflecting elements are passive and that no power is needed when it reflects the incident signal\footnote{The consideration of passive reflecting elements is suitable when we deploy a large number of reflecting elements at THz bands while active IRS is suitable for a small number of reflecting elements~\cite{long2021active}. Moreover, some switch circuits are used at passive IRS, enabling each reflecting element to reconfigure its reflection coefficient, which requires minimal power, i.e., tens of microwatt per reflecting element~\cite{huang2019reconfigurable}. }~\cite{long2021active}. The set of ${N}$ available distributed IRSs is denoted as $\mathcal{I} = \{\IRS_1, \IRS_2,\ldots, \IRS_{{N}} \}$, where the $n^{th}$ IRS is equipped with $M_n$ reflecting elements. The area of each IRS element is $A = M_x \times M_y$, where $M_x$ and $M_y$ denote the length of the horizontal and vertical sides, respectively. Henceforth, we use T, R, and I as acronyms for the Tx, Rx, and IRS, respectively. {In addition, we assume all direct links are blocked\footnote{{This assumption is reasonable when the direct link is severely obstructed by obstacles or human bodies in the environment~\cite{wan2021terahertz,fang2022optimum}.}} due to severe shadowing by obstacles or human bodies in the environment~\cite{wan2021terahertz,fang2022optimum}, and receivers are served through IRS cascaded links, as shown in Fig.~\ref{m1}.} Through a well-planned deployment of the IRSs, we consider that there exists a LOS path between the $\tx$-to-$\IRS$ link and the $\IRS$-to-$\rx$ link. 
In THz channels, the pathloss of the non-LOS (NLOS) links is known to be much larger than that of the LOS links due to
smaller wavelengths~\cite{khawaja2020coverage}. Therefore, we ignore the NLOS links between the $\tx$-to-$\IRS$ link and the $\IRS$-to-$\rx$ link. Furthermore, we only consider signals reflected by an IRS one time because the signals reflected by the IRS two times or more are weak and can be neglected~\cite{noh2022cell}.
\subsection{3D Cartesian Coordinates $(x, y, z)$}
In the Cartesian coordinate system, the center locations of the $n^{th}$ IRS ($\IRS_n$), the $k^{th}$ $\tx$ ($\tx_k$), and the $l^{th}$ $\rx$ ($\rx_l$) are defined as  $\vec c_{\IRS}^n = (x_{\IRS}^n, y_{\IRS}^n, z_{\IRS}^n)$, $\vec c_{\tx}^k = (x_{\tx}^k, y_{\tx}^k, z_{\tx}^k)$, and $\vec c_{\rx}^l = (x_{\rx}^l, y_{\rx}^l, z_{\rx}^l)$, respectively. The vectors joining $\tx_k$ to the $m^{th}$ reflecting element on $\IRS_n$ ($\IRS_{n,m}$) and $\IRS_{n,m}$-to-$\rx_l$ are expressed as $\vec{r}_{k,n,m} =(x_{\tx}^k-x_{\IRS}^{n,m}, y_{\tx}^k-y_{\IRS}^{n,m}, z_{\tx}^k-z_{\IRS}^{n,m}),$ and  $\vec{r}_{n,m,l} =(x_{\rx}^k-x_{\IRS}^{n,m}, y_{\rx}^k-y_{\IRS}^{n,m}, z_{\rx}^k-z_{\IRS}^{n,m}),$ respectively.
\subsection{ 3D Cartesian to Spherical Coordinates (d, $\psi$, $\phi$)}
\begin{figure}[!t]
 \centering
 \includegraphics[width=\linewidth]{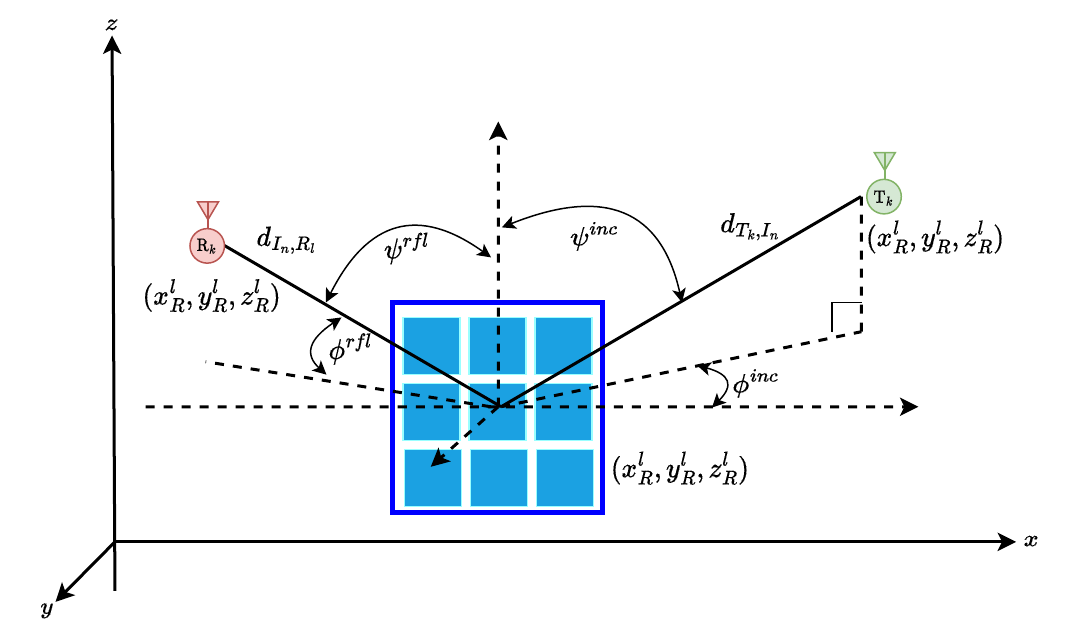}
 \caption{Illustration of 3D geometry of the  IRS-aided network considered in the channel model.
 }
 \label{m2}
 \end{figure}

The Cartesian coordinates $(x,y,z)$ of a node's location can be converted into spherical coordinates ($d$, $\psi$, $\phi$), where $d$, $\psi$, and $\phi$ represent the distance, elevation angle, and azimuth angle, respectively. Let $d_{\tx_k,\IRS_{n,m}}$ denote the distance from $\tx_k$ to $\IRS_{n,m}$, and $d_{\IRS_{n,m},\rx_l}$ denote the distance from $\IRS_{n,m}$ to $\rx_l$, as shown in Fig.~\ref{m2}. For THz communications, the elements of the IRS are densely installed, and we ignore the distance between the IRS elements. Thus, the distances $d_{\tx_k,\IRS_n}$ and $d_{\IRS_n,\rx_l}$ can be computed from the center location of $\IRS_n$ to $\tx_k$ and $\rx_l$, respectively, which are expressed as 
 \begin{align}
  d_{\tx_k,\IRS_n} &=\textstyle \sqrt{(x_{\tx}^k-x_{\IRS}^n)^2 + (y_{\tx}^k-y_{\IRS}^n)^2 + (z_{\tx}^k-z_{\IRS}^n)^2},
 \end{align}
 \begin{align}
 d_{\IRS_n,\rx_l} & = \textstyle\sqrt{(x_{\rx}^k-x_{\IRS}^n)^2 + (y_{\rx}^k-y_{\IRS}^n)^2 + (z_{\rx}^k-z_{\IRS}^n)^2}.
 \end{align}
\textit{Elevation Angle:} In order to determine the height of the node, the angle between the $z$-axis and the radius is measured, known as the elevation angle. The unit vectors in the direction of the $x$-axis, $y$-axis, and $z$-axis are given by $\vec{r}_{x} =(1,0,0)$, $ \vec{r}_{y} =(0,1,0)$, and 
$ \vec{r}_{z} =(0,0,1)$, respectively.
The cosine of $\psi^\icd_{k,n,m}$ is obtained~as
\begin{align}
    \cos{(\psi^\icd_{k,n,m})} &=\bigg[{\frac{\vec{r}_{k,n,m} \vec{r}_{z}}{\Vert\vec{r}_{k,n,m}\Vert \hspace{3pt} \Vert\vec{r}_{z}\Vert}} \bigg] = \bigg [{\frac{ z_\tx^k - z_{\IRS}^{n,m} }{d_{\tx_k,\IRS_{n,m}}}} \bigg]. 
\end{align}

As a result, the cosine of $\psi^\rfl_{n,m,l}$ from $\IRS_{n, m}$ to $\rx_l$ is calculated~as 
\begin{align}
    \cos{(\psi^\rfl_{n,m,l})} = \bigg[\frac{z_\rx^l - z_{\IRS}^{n,m} }{ d_{\IRS_{n,m},\rx_l}}\bigg].
\end{align}
\textit{Azimuth Angle:} 
The azimuth angle is measured in relation to the $x$-axis and to the radius projected into the 2D plane, which determines the direction of the node. The tangent of $\phi^\icd_{k,n,m}$ can be written as
\begin{align}
    \tan{(\phi^\icd_{k,n,m})} &=\bigg[{\frac{\vec{r}_{k,n,m} \vec{r}_{y}}{\vec{r}_{k,n,m} \vec{r}_{x}}} \bigg] = \bigg [{\frac{ y_\tx^k - y_{\IRS}^{n,m}}{ x_\tx^k - x_{\IRS}^{n,m}}} \bigg]. 
\end{align}

As a result, the tangent of $\phi^\rfl_{n,m,l}$ from $\IRS_{n,m}$ to $\rx_l$ is obtained~as 
\begin{align}
    \tan{(\phi^\rfl_{n,m,l})} = \bigg[ \frac{ y_\rx^l - y_{\IRS }^{n,m}} {x_\rx^l - x_{\IRS}^{n,m} }\bigg].
\end{align}
\subsection{IRS Phase-Shift Configurations}
Introducing a PS to impinging signals is an important characteristic of the IRS. The PS matrix combines the amplitude and PS of each element of the IRS. The PS matrix of the $\IRS_n$ can be described as~\cite{Bjornson_WCL_2020}: 
\begin{align}
    \vec{\Theta}_n = \mathrm{diag}([\kappa_{n,1} e^{j \theta_{n,1}}, ..., \kappa_{n,m} e^{j \theta_{n,m}}, ..., \kappa_{n,M} e^{j \theta_{n,M}}]),
\end{align}
where $\kappa_{n,m} \in (0,1]$ and $\theta_{n,m} \in [0, 2\pi)$ represent the amplitude and the PS of $\IRS_{n,m}$, respectively. The PS matrix can be categorized into three categories depending on the deployment of the IRS \cite{liu2021reconfigurable}, namely ($i$) continuous amplitude and PS, where the amplitude and PS of the IRS elements are continuously adjusting, ($ii$) constant amplitude and continuous PS, where the amplitude is constant, e.g., $\kappa_{n,m} = 1$, and the PS can be continuously adjusted, and ($iii$) constant amplitude and discrete PS, where the amplitude is fixed and the PS is adjusted based on a discrete set of values.
\subsection{Channel Modeling of the Considered IRS-THz System}
Assuming that $\tx_k$ transmits a signal $x_k$ with transmission power $p_{k}$, where ${\mathbb{E}}[ |x_k|^2]= 1$. We assume that the direct link between the $\tx$ and the $\rx$ is blocked by an obstacle, and therefore focus on the propagation model for cascaded links. {Additionally, we consider that the CSI available at the transmitters, receivers, and IRSs is imperfect\footnote{Channel estimation is much more difficult in IRS-assisted wireless communication than in traditional wireless networks. Furthermore, it is more challenging at the IRS than the Tx / Rx because the IRS has passive reflecting elements that cannot process the pilot signals sent from and to the Tx / Rx~\cite{pan2021reconfigurable, shah2022statistical}. However, channel reciprocity holds for the Tx-to-IRS and IRS-to-Rx channels~\cite{tang2021channel}. We, therefore, assume that the estimated CSI at the Tx / Rx is similar to the IRS's CSI.}}.

The wireless channel can be achieved by either the near-field or far-field models. The communications in the near-field regions are spherical-wave-based, while planar-wave-based in the far-field regions. To differentiate the near-field or far-field regions, we define the Rayleigh distance $d_{ray}$ as a boundary, which can be expressed as \cite{Cui_MCOM_2023}
\begin{align}\label{rray}
    d_{ray}=\frac{2 D^2}{\lambda}.
\end{align}
where $D$ is the aperture of the IRS. Moreover, the range of near-field region can be expressed as~\cite{Cui_MCOM_2023}
\begin{align}\label{range_NFC}
     d_{\tx_k,\IRS_n} < d_{ray},\quad d_{\IRS_n,\rx_l} < d_{ray}.
\end{align}
The range after the near-field region in~\eqref{range_NFC} is considered as the far-field region, and the communication in this region is far-field communication. For the carrier frequency $f_c=300$ GHz, $\lambda=0.001\, m$, and side length of IRS elements  $0.4 \lambda$~\cite{najafi2020physics}, we calculate the Rayleigh distance $d_{ray}$ for different numbers of IRS elements as shown in Table~\ref{tab:far_near}.
\begin{table}[!htp]
\centering
\renewcommand{\arraystretch}{1}
\caption{Rayleigh distance $d_{ray}$ with varying number of IRS elements.}
\label{tab:far_near}
\begin{tabular}{l l l}
\hline
{$M=M_x\times M_y$}                                  & Physical size of IRS [$m^2$] & $d_{ray}\, [m]$ \\ \hline
$30\times30$&$0.012\times 0.012$& 0.288\\
$50\times50$&$0.02\times 0.02$&0.8\\
$100\times100$&$0.04\times 0.04$&3.2      
\\ \hline
\end{tabular}
\end{table}
From Table~\ref{tab:far_near}, we verify that the Rayleigh distance $d_{ray}$ is very small which makes the range of near-field communication is very limited. Thus we ignore this region and consider the far-field communication.
\subsubsection{Tx to IRS Channel Modeling}
Let $\widehat{\vec{h}}_{\tx_k,\IRS_n}{ \in \mathbb{C}}^{M \times 1}$ denote the estimated channel (EC) vector from $\tx_k$-to-$\IRS_n$. In this context, $[\widehat{\vec{h}}_{\tx_k,\IRS_n}]_m = \widehat{h}_{\tx_k,\IRS_{n,m}}$ is the individual channel from $\tx_k$-to-$\IRS_{n,m}$ which can be expressed as~\cite{DovelosICC2021}
\begin{align}
    [\widehat{\vec{h}}_{\tx_k,\IRS_n}]_m = \widehat{h}_{\tx_k,\IRS_{n,m}} = \sqrt{\pl_{\tx_k,\IRS_{n,m}}} e^{ - j \frac{2 \pi}{\lambda} d_{\tx_k,\IRS_{n,m}}},
\end{align}
where
$\frac{2 \pi}{\lambda}$ is the wavenumber and $\pl_{\tx_k,\IRS_{n,m}}$ represents the PL of the $\tx_k$-to-$\IRS_{n,m}$ link which can be expressed as~\cite{OzcanTVT2021,ozdogan2019intelligent}
\begin{align} \label{plsr}
    \pl_{\tx_k,\IRS_{n,m}}=\gain_{\tx_k} \gain_{\IRS_{n,m}}(-\vec{r}_{k,n,m})\frac{A^2 e^{-\kappa_{abs}(f)d_{\tx_k,\IRS_{n,m}}}}{(4 \pi d_{\tx_k,\IRS_{n,m}})^2},
\end{align}
where $\kappa_{abs}(f)$ is the absorption coefficient at the carrier frequency $f$, $\gain_{\tx_k}$ is antenna gain at $\tx_k$, and $\gain_{\IRS_{n,m}}(-\vec{r}_{k,n,m})$ denotes the gains of $\IRS_{n,m}$ in the incident direction, which can be written as \cite{hu2021angle}
\begin{align}
  \vec{r}_{k,n,m} =  d_{\tx_k,\IRS_{n,m}}\begin{pmatrix} \cos{(\psi^\icd_{k,n,m})} \cos{(\phi^\icd_{k,n,m})} \\ \cos{(\psi^\icd_{k,n,m})} \sin{(\phi^\icd_{k,n,m})} \\ \sin{(\psi^\icd_{k,n,m})}\end{pmatrix} ,
\end{align}
where $\psi$ and $\phi$ are the elevation and azimuth angles
from the IRS unit to a certain transmitting/receiving direction. Furthermore, the actual channel with the CSI estimation errors (CEEs) can be expressed as 
\begin{align} \vec{h}_{\tx_k,\IRS_n}=\widehat{\vec{h}}_{\tx_k,\IRS_n}+\widetilde{\vec{h}}_{\tx_k,\IRS_n},
\end{align}
where $\widetilde{\vec{h}}_{\tx_k,\IRS_n}$ denotes the CEEs, which are uncorrelated with the estimated channel $\widehat{\vec{h}}_{\tx_k,\IRS_n}$, and the entries of $\widetilde{\vec{h}}_{\tx_k,\IRS_n}$ are IID complex Gaussian with zero mean and variance $\sigma_{\widetilde{\vec{g}}_{\tx_k,\IRS_n}}^2$.

\subsubsection{IRS to Rx Channel Modeling}
The received power $p_{n}~=~p_{k}\pl_{\tx_k,\IRS_{n,m}}$ at $\IRS_n$ from $\tx_k$ can be calculated using the PL in~\eqref{plsr}. Let $\widehat{\vec{g}}_{\IRS_n,\rx_l} \in \mathbb{C}^{M \times 1}$ denotes the EC gain vector of the $\IRS_n$-to-$\rx_l$ links, and $[\widehat{\vec{g}}_{\IRS_n,\rx_l}]_m = \widehat{g}_{\IRS_{n,m},\rx_l}$ be the individual channel from $\IRS_{n,m}$-to-$\rx_l$, which can be expressed as~\cite{DovelosICC2021}
\begin{align}
   [\vec{g}_{\IRS_n,\rx_l}]_m = g_{\IRS_{n,m},\rx_l} = \sqrt{\pl_{\IRS_{n,m},\rx_l}} e^{ - j \frac{2 \pi}{\lambda} d_{\IRS_{n,m},\rx_l}} ,
\end{align}
where $\pl_{\IRS_{n,m},\rx_l}$ represents the PL of the $\IRS_{n,m}$-to-$\rx_l$ link which can be expressed as
\begin{align} \label{plru}
    \pl_{\IRS_{n,m},\rx_l}&=\gain_{\IRS_{n,m}}(\vec{r}_{n,m,l}) \gain_{\rx_l} \frac{A^2 e^{-\kappa_{abs}(f)d_{\IRS_{n,m},\rx_l}}}{(4 \pi d_{\IRS_{n,m},\rx_l})^2},
\end{align}
where $\gain_{\rx_l}$ is antenna gain at $\rx_l$ and $\gain_{\IRS_{n,m}}(\vec{r}_{n,m,l})$ are the gains of $\IRS_{n,m}$ in the reflection direction $\vec{r}_{n,m,l}$, which can be written as \cite{hu2021angle}
\begin{align}
  \vec{r}_{n,m,l} =   d_{\IRS_{n,m},\rx_l}\begin{pmatrix} \cos{(\psi^\rfl_{n,m,l})} \cos{(\phi^\rfl_{n,m,l})} \\ \cos{(\psi^\rfl_{n,m,l})} \sin{(\phi^\rfl_{n,m,l})} \\ \sin{(\psi^\rfl_{n,m,l})}\end{pmatrix}.
\end{align}
Meanwhile, the channel with CEEs $\widetilde{\vec{g}}_{\IRS_n,\rx_l}\sim \mathcal{CN}(\mathbf{0}, \sigma^2_{\widetilde{\vec{g}}_{\IRS_n,\rx_l}})$ can be expressed as
\begin{align}
{\vec{g}}_{\IRS_n,\rx_l}=\widehat{\vec{g}}_{\IRS_n,\rx_l}+\widetilde{\vec{g}}_{\IRS_n,\rx_l}.
\end{align}
The reflecting elements of the IRS can be designed to be identical in size and gain, with small size and low gain in the THz band. Therefore, the reflecting elements can have the same normalized power radiation pattern model given as~\cite{nguyen2022channel}
\begin{align}\label{rad}
\mathbb{U}_{\IRS_{n,m}}(\psi)=\cos^q(\psi),\quad 0\leq\,\psi\,<{\pi/2},
\end{align}
where $q$ determines the directivity of each reflecting element. the above pattern solely relies on the elevation angle $\psi$, where the maximum antenna gain towards the IRS can be obtained with zero elevation angle, i.e., $\psi=0$~\cite{ellingson2021path}. Furthermore, the directivity of the reflecting element is also a function of the elevation angle $\psi$ and is given as
\begin{align}
    \mathbb{D}_{\IRS_{n,m}}(\psi)=2(q+1)\cos^q(\psi),\quad 0\leq\,\psi\,<{\pi/2}.
\end{align}
The reflecting element's directivity is governed by the value of $q$ in this model. In addition, the pattern in~\eqref{rad} is a cosine function raised to the power $q$, and as the value of $q$ increases, the beam becomes more directive. Thus the gain of $\IRS_{n,m}$ can be expressed as
\begin{align} \label{eq:GE}
    \gain_{\IRS_n}(\psi, \phi)=\begin{cases} 2\gamma(q+1) \cos^{q}(\psi), &	0\leq\,\psi\,<{\pi/2}, \\
    0,      	& {\pi/2}\leq\,\psi\,\leq{\pi},
    \end{cases}
\end{align}
where $\phi \in[0, 2 \pi )$ and $\gamma$ is the antenna efficiency of the IRS reflecting elements. Thus, the existing model is only applicable for a specific value of $q$ and zero elevation angle, i.e., $\psi=0$. Therefore, based on the existing PL models for THz frequencies~\cite{DovelosICC2021, OzcanTVT2021}, we propose a new PL model that facilitates our network analysis.
\subsubsection{Cascaded Channel Modeling}
IRS elements scatter the incident wave in all directions rather than reflecting it with the same beam curvature. Therefore, some signals can be lost, which causes extra PL\footnote{The amplitude of the IRS elements $\kappa_{n,m}\in(0,1]$, when the amplitude of the IRS element is less than 1, it means that the IRS is not fully reflecting the incident signal's power. In other words, some of the signal power is lost when it interacts with the IRS. Therefore, we set $\kappa_{n,m}=1$ to avoid the extra PL~\cite{ni2021resource,sun2023irs}.}. In the case of IRS-aided networks, the received signal power depends on the area and effective aperture of the IRS. When the IRS elements become infinitely large, they act as perfect mirrors that reflect all the incident waves according to Snell's reflection law~\cite{balanis2012advanced}. Let $h_{\tx_k,\IRS_n,\rx_l}$ be the cascaded channel from $\tx_k$-to-$\rx_l$ through $\IRS_n$, which can be expressed as
\begin{align}
    &h_{\tx_k,\IRS_n,\rx_l}=(\vec{h}_{\tx_k,\IRS_n})^T \vec{\Theta}_n \vec{g}_{\IRS_n,\rx_l}\nonumber\\&=(\widehat{\vec{h}}_{\tx_k,\IRS_n}+\widetilde{\vec{h}}_{\tx_k,\IRS_n})^T \vec{\Theta}_n (\widehat{\vec{g}}_{\IRS_n,\rx_l} +\widetilde{\vec{g}}_{\IRS_n,\rx_l} )\nonumber\\&=\underbrace{\widehat{\vec{h}}_{\tx_k,\IRS_n}^T\vec{\Theta}_n \widehat{\vec{g}}_{\IRS_n,\rx_l}}_{\widehat{h}_{\tx_k,\IRS_n,\rx_l}} \nonumber\\&+\underbrace{\widetilde{\vec{h}}_{\tx_k,\IRS_n}^T \vec{\Theta}_n \widehat{\vec{g}}_{\IRS_n,\rx_l}+\widehat{\vec{h}}_{\rx_k,\IRS_n}^T\vec{\Theta}_n \widetilde{\vec{g}}_{\IRS_n,\rx_l} \!+\!\widetilde{\vec{h}}_{\tx_k,\IRS_n}^T \vec{\Theta}_n \widetilde{\vec{g}}_{\IRS_n,\rx_l}}_{\widetilde{h}_{\tx_k,\IRS_n,\rx_l}},
\end{align}
where $\widetilde{{h}}_{\tx_k,\IRS_n,\rx_l}$ represents the CEEs, which is modeled as 
$\mathcal{CN}(\mathbf{0},\sigma_{\widetilde{g}_{\IRS_n,\rx_l}}^2|\vec{\widehat{h}}^T_{\tx_k,\IRS_n}\vec{\widehat{h}}_{\tx_k,\IRS_n}|+\sigma_{\widetilde{h}_{\tx_k,\IRS_n}}^2|\vec{\widehat{g}}^T_{\IRS_n,\rx_l}\vec{\widehat{g}}_{\IRS_n,\rx_l}|+\sigma_{\widetilde{h}_{\tx_k,\IRS_n}}^2\sigma_{\widetilde{g}_{\IRS_n,\rx_l}}^2)$.
\begin{Lemma} \label{lemma_gain}
Denoting ${ \gain_{\IRS_{n,m}}(\vec{r}^{\icd}, \vec{r}^{\rfl}) \triangleq  \gain_{\IRS_{n,m}}(-\vec{r}^{\icd}) \gain_{\IRS_{n,m}}(\vec{r}^{\rfl}) }$ as the cascaded gain of the $m^{th}$ element of $\IRS_n$, the angle-based trigonometric model of ${\gain_{\IRS_{n,m}}(\vec{r}^{\icd}, \vec{r}^{\rfl})}$ is formulated as 
\begin{align} \label{gain}
    \gain_{\IRS_{n,m}}(\vec{r}^{\icd}, \vec{r}^{\rfl}) 
&=  \bigg( 
        \frac{4 \pi A}{\lambda^2} 
        \eta_{\IRS_{n,m}}(-\vec{r}^{\icd}, \vec{r}^{\rfl})
    \bigg)^2, \\
\eta_{\IRS_{n,m}}(-\vec{r}^{\icd}, \vec{r}^{\rfl})
&\triangleq
    \sqrt{ \cos^2 \psi^{\icd} (\cos^2 \phi^{\rfl} \cos^2 \psi^{\rfl} +\sin^2 \phi^{\rfl})}. \nonumber
\end{align}
\end{Lemma}
\begin{IEEEproof}
The proof is provided in Appendix~\ref{ap_lemma1}.
\end{IEEEproof}
 \begin{figure}[!t]
 \centering
 \includegraphics[width=\linewidth]{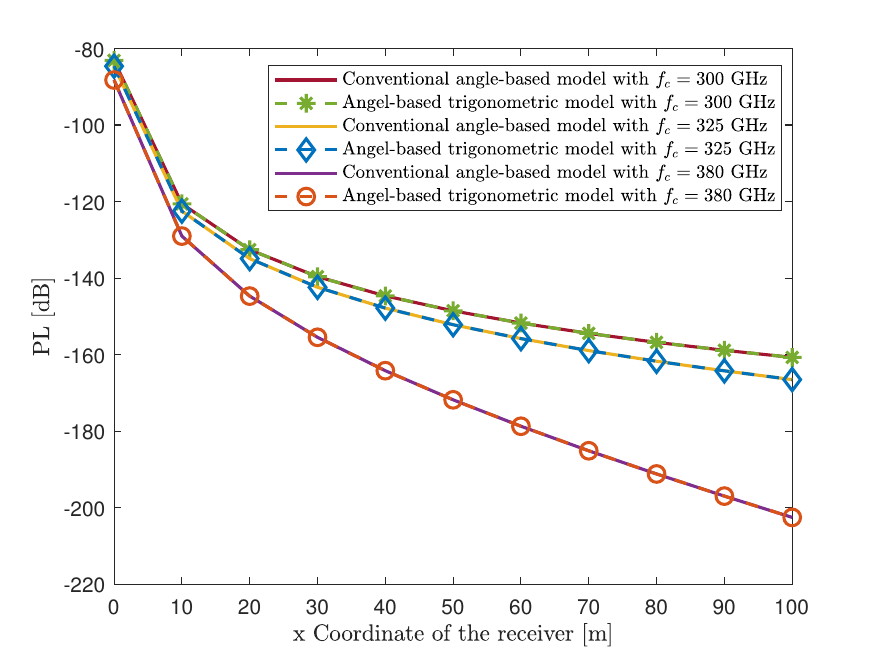}
 \caption{Comparison between the conventional angel-based and the proposed angel-based trigonometric PL models for different frequencies.}
 \label{pl_difmodel}
 \end{figure}
Invoking Lemma~\ref{lemma_gain}, the PL of the cascaded link is the combination of the PLs over the $\tx_k$-to-$\IRS_n$ and $\IRS_n$-to-$\rx_l$ links, which can be expressed as
\begin{align} \label{pldr}
    \pl_{\tx_k,\IRS_{n},\rx_l} &= \gain_{\tx_k}  \gain_{\IRS_{n,m}}(-\vec{r}_{k,n,m}) \gain_{\IRS_{n,m}}(\vec{r}_{n,m,l}) \gain_{\rx_l}\nonumber\\&\quad \times \frac{ \lambda^4 e^{-\kappa_{abs}(f)(d_{\tx_k,\IRS_{n,m}}+d_{\IRS_{n,m},\rx_l})}}{(4 \pi)^4 (d_{\tx_k,\IRS_{n,m}} d_{\IRS_{n,m},\rx_l})^2}.
\end{align}
The PL, assuming the conventional angle-based and the proposed angle-based trigonometric models with a single Tx/Rx, is demonstrated in Fig.~\ref{pl_difmodel} for different frequencies.
\subsection{$\SINR$ Formulation with Arbitrary Phase Shifts}
Next, we present the expression of T-I-R $\SINR$ for an IRS-aided network over the THz band. Assuming that $x_k (t)$ is the transmit signal of $\tx_k$ at time slot $t$, the received signal $y_{\tx_k,\IRS_{n},\rx_l} (t)$ at $\rx_l$ is given in~\eqref{eq_rx_signal}, at the top of next page,
\begin{figure*}[!htp]
	\begin{align} \label{eq_rx_signal}
		 y_{\tx_k,\IRS_{n},\rx_l} (t) &= \textstyle\sqrt{p_{k}}\widehat{h}_{\tx_k,\IRS_n,\rx_l} x_k (t)
		+ \sum\limits_{\substack{j=1, j\neq k}}^K \sum\limits_{\substack{i=1 }}^N\sqrt{p_{j}}\widehat{h}_{\tx_j,\IRS_i,\rx_l} x_j (t)+ \sum\limits_{\substack{j=1}}^K \sum\limits_{\substack{i=1 }}^N\sqrt{p_{j}}\widetilde{h}_{\tx_j,\IRS_i,\rx_l} x_j (t)
		+ w_l(t)
\nonumber\\
  &=\textstyle\underbrace{\sqrt{p_{k}}(\widehat{\vec{h}}_{\tx_k,\IRS_n})^T \vec{\Theta}_n \widehat{\vec{g}}_{\IRS_n,\rx_l} x_k (t)}_\text{desired signal}
		+ \underbrace{\sum\limits_{\substack{j=1, j\neq k}}^K \sum\limits_{\substack{i=1 }}^N\sqrt{p_{j}}(\widehat{\vec{h}}_{\tx_j,\IRS_i})^T \vec{\Theta}_i \widehat{\vec{g}}_{\IRS_i,\rx_l} x_j (t)}_\text{interference from another transmitters} \nonumber\\
  &+ \underbrace{\sum\limits_{\substack{j=1}}^K \sum\limits_{\substack{i=1 }}^N\sqrt{p_{j}}\big(\widetilde{\vec{h}}_{\tx_j,\IRS_i}^T \vec{\Theta}_i \widehat{\vec{g}}_{\IRS_i,\rx_l}+\widehat{\vec{h}}_{\tx_j,\IRS_i}^T\vec{\Theta}_i \widetilde{\vec{g}}_{\IRS_i,\rx_l} +\widetilde{\vec{h}}_{\tx_j,\IRS_i}^T \vec{\Theta}_i \widetilde{\vec{g}}_{\IRS_i,\rx_l}\big) x_j (t)}_\text{interference due to imperfect CSI}
		+ \underbrace{w_l(t)}_\text{noise},
	\end{align}
 \hrule
 \end{figure*}
\begin{figure*}[!htp]
	\begin{align} \label{eq_sinr_e2e1}
		&\SINR_{\tx_k,\IRS_n,\rx_l}=\frac{p_k|\widehat{h}_{\tx_k,\IRS_n,\rx_l}|^2}{\sum\limits_{\substack{j=1, j\neq k}}^K \sum\limits_{\substack{i=1}}^N p_j\big|\widehat{h}_{\tx_j,\IRS_i,\rx_l}\big|^2+\sum\limits_{\substack{j=1}}^K \sum\limits_{\substack{i=1}}^N p_j\big|\widetilde{h}_{\tx_j,\IRS_i,\rx_l}\big|^2+\sigma_l^2}\nonumber\\&=\frac{p_k\big|(\widehat{\vec{h}}_{\tx_k,\IRS_n})^T\vec{\Theta}_n\widehat{\vec{g}}_{\IRS_n,\rx_l}\big|^2}{\sum\limits_{\substack{j=1, j\neq k}}^K \sum\limits_{\substack{i=1}}^N p_j\big|(\widehat{\vec{h}}_{\tx_j,\IRS_i})^T\vec{\Theta}_i\widehat{\vec{g}}_{\IRS_i,\rx_l}\big|^2+\sum\limits_{\substack{j=1}}^K \sum\limits_{\substack{i=1}}^N p_j\big(\sigma_{\widetilde{g}_{\IRS_i,\rx_l}}^2\big|\vec{\widehat{h}}^T_{\tx_j,\IRS_i}\vec{\widehat{h}}_{\tx_j,\IRS_i}\big|+\sigma_{\widetilde{h}_{\tx_j,\IRS_i}}^2\big|\vec{\widehat{g}}^T_{\IRS_i,\rx_l}\vec{\widehat{g}}_{\IRS_i,\rx_l}\big|+\sigma_{\widetilde{h}_{\tx_j,\IRS_i}}^2\sigma_{\widetilde{g}_{\IRS_i,\rx_l}}^2\big)+\sigma_l^2}\nonumber\\&=\frac{\frac{p_k\lambda^2}{16\pi^2}\Big|\sum\limits_{m=1}^M \frac{ \sqrt{ \gain_{\tx_k} \gain_{\rx_l} \gain_{\IRS_{n,m}}(-\vec{r}_{k,n,m})\gain_{\IRS_{n,m}}(\vec{r}_{n,m,l}) e^{-\kappa_{abs}(f)(d_{\tx_k,\IRS_{n,m}}+d_{\IRS_{n,m},\rx_l})}}}{ d_{\tx_k,\IRS_{n,m}} d_{\IRS_{n,m},\rx_l}}\kappa_{n,m} e^{ - j \frac{2 \pi}{\lambda} (d_{\tx_k,\IRS_{n,m}}+d_{\IRS_{n,m},\rx_l}){+j \theta_{n,m}} }
			\Big|^2}{\sum\limits_{\substack{j=1, j\neq k}}^K \sum\limits_{\substack{i=1}}^N p_j\big|(\widehat{\vec{h}}_{\tx_j,\IRS_i})^T\vec{\Theta}_i\widehat{\vec{g}}_{\IRS_i,\rx_l}\big|^2+\sum\limits_{\substack{j=1}}^K \sum\limits_{\substack{i=1}}^N p_j\big(\sigma_{\widetilde{g}_{\IRS_i,\rx_l}}^2\big|\vec{\widehat{h}}^T_{\tx_j,\IRS_i}\vec{\widehat{h}}_{\tx_j,\IRS_i}\big|+\sigma_{\widetilde{h}_{\tx_j,\IRS_i}}^2\big|\vec{\widehat{g}}^T_{\IRS_i,\rx_l}\vec{\widehat{g}}_{\IRS_i,\rx_l}\big|+\sigma_{\widetilde{h}_{\tx_j,\IRS_i}}^2\sigma_{\widetilde{g}_{\IRS_i,\rx_l}}^2\big)+\sigma_l^2},
	\end{align}
 \hrule
 \end{figure*}
 where the first part is the desired signal; interfering signals are represented in the second part; $w_l(t)$ denotes the additive white Gaussian noise (AWGN) at the user for the cascaded channel; $[\vec{\Theta}_n]_m = \kappa_{n,m} e^{ j\theta_{n,m} }$ is the $m^{th}$ element on the diagonal that runs from the top left to the bottom right of matrix $\vec{\Theta}_n$. Thus, the SINR at $\rx_l$, $\SINR_{\tx_k\IRS_n\rx_l}$, of the desired signal from $\tx_k$-to-$\IRS_n$-to-$\rx_l$ link is given in~\eqref{eq_sinr_e2e1}, where $\sigma_l^2$ is the AWGN noise power at each receiver.
\section{Problem Formulation}\label{pform}
As stated earlier, our primary goal is T-I-R scheduling to maximize the sum rate. For our system model, the achievable rate can be written as
\begin{align}\label{rateknl}
     R_{\tx_k,\IRS_n,\rx_l} ([\vec{\Omega}]_{\tx_k,\IRS_n,\rx_l})= \log_2(1+\SINR_{\tx_k,\IRS_n,\rx_l} ([\vec{\Omega}]_{\tx_k,\IRS_n,\rx_l})).
\end{align}
Considering the $\SINR$ as a function of $[\vec{\Omega}]_{\tx_k,\IRS_n,\rx_l}$,
where $[\vec{\Omega}]_{\tx_k,\IRS_n,\rx_l}$ is the allocation matrix with each element of allocation matrix being a 3-dimension, i.e., $(\tx_k,\IRS_n,\rx_l)$. Therefore, the objective is to find the optimal allocation matrix $[\vec{\Omega}]_{\tx_k,\IRS_n,\rx_l}$ that maximizes the sum rate, expressed as
\begin{subequations}\label{eq_opt_prob}
\begin{alignat}{2}
&\underset{[\vec{\Omega}]_{\tx_k,\IRS_n,\rx_l},{\vec{\Theta}_n}}{\mathrm{maximize}}
&\quad
&\textstyle  \sum_{l=1}^{L}  R_{\tx_k,\IRS_n,\rx_l} ([\vec{\Omega}]_{\tx_k,\IRS_n,\rx_l}), \label{eq_optProb}\\
&\text{subject to} 
&&\textstyle  \sum_{k=1}^K [\vec{\Omega}]_{\tx_k,\IRS_n,\rx_l} = 1 , \forall{n,l}, \label{eq_constraint1}\\
&&&\textstyle  \sum_{n=1}^N [\vec{\Omega}]_{\tx_k,\IRS_n,\rx_l} = 1  , \forall{k,l}, \label{eq_constraint2}\\
&&& \textstyle  \sum_{l=1}^{L} [\vec{\Omega}]_{\tx_k,\IRS_n,\rx_l} = 1  , \forall{k,n}, \label{eq_constraint3}\\
&&&  [\vec{\Omega}]_{\tx_k,\IRS_n,\rx_l} \in \{0,1\}, \forall{k,n,l}, \label{eq_constraint4}\\
&&&  {\theta_{n,m}} \in [0,2\pi), \kappa_{n,m} \leq 1, \forall{n,m}, \label{eq_constraint7}
\end{alignat}
\end{subequations}
where the objective \eqref{eq_optProb} is the sum rate of the network given the pairing $[\vec{\Omega}]_{\tx_k,\IRS_n,\rx_l}$ with IRS reflection $\vec{\Theta}_n$; constraint~\eqref{eq_constraint1} ensures that each T will be paired to one I and R only; constraint~\eqref{eq_constraint2} guarantees that each I is allocated to one T and R; constraint~\eqref{eq_constraint3} make sure that each R is paired with one I and T. Thus, constraint~\eqref{eq_constraint1}, \eqref{eq_constraint2}, and \eqref{eq_constraint3} ensure a one-to-one T-I-R matching\footnote{We want to serve all T/R pairs, therefore, we allocate one IRS to assist the communication for one pair (T/R). Furthermore, multiple IRS allocations for one link cause a strong interference with the other links~\cite{mei2021performance}. Thus, we allocate one IRS to serve one link in the network.}. Additionally, constraint~\eqref{eq_constraint4} enforces that $[\vec{\Omega}]_{\tx_k,\IRS_n,\rx_l}$ is either 1 i.e., the IRS is allocated, or 0 i.e., the IRS is free and~\eqref{eq_constraint7} restricts the PS and amplitude of the reflection model. 

The formulated problem is a MINLP problem because it includes multiple integer variables, e.g., $[\vec{\Omega}]_{\tx_k,\IRS_n,\rx_l} \in \{0,1\}$, and continuous variables $\theta_{n,m} \in [0, 2\pi]$, and the objective function is nonlinear~\cite{cui2017optimal,alizadeh2019load,sun2021sum}. Additionally, it can be observed that the optimization problem in~\eqref{eq_opt_prob} includes the optimization of the IRSs' reflection and a 3D matching problem involving three disjoint sets (i.e., Txs, IRSs, and Rxs). Thus, our MINLP problem in~\eqref{eq_opt_prob} is NP-hard since it combines optimizing over discrete variables with a challenge of dealing with nonlinear functions, and therefore its solution usually involves an enormous search space with exponential time complexity~\cite{jung2013perspective}.
\section{Optimal Solution and Problem Reformulation}\label{optimal_solution}
With NP-hard problems, compute-intensive ES is conventionally used for finding the optimal solution~\cite{burer2012non}. As can be seen in \eqref{eq_opt_prob}, $\theta_{n,m}$ and $[\vec{\Omega}]_{\tx_k,\IRS_n,\rx_l}$ are functions of $(\tx_k,\IRS_n,\rx_l)$. Assuming that the imperfect CSI is available, we can obtain the optimal PS for any pairing, as explained next in \eqref{eq_ps_opt}. In this context, optimizing the PS and pairing are two independent problems. Thus, to address the optimization problem in~\eqref{eq_opt_prob}, we search over all possible pairings, each pairing uses an optimal PS configuration to achieve the best potential performance of that pair. As a result, the obtained result is the optimal solution to problem \eqref{eq_opt_prob}. In what follows, we present the two steps of the optimal solution: (i) setting the optimal PS configuration for an arbitrary pair, i.e., a cascaded transmission from a $\tx$-to-$\IRS$-to-$\rx$, and (ii) finding the optimal $\tx$-$\IRS$-$\rx$ parings for the entire network.
\subsection{Phase Shift Optimization} \label{optimalref}
Here, we optimize the reflection model to maximize the received signal strength by constructive addition of multiple reflected signals. In order to maximize the power gain, the PS induced by $\IRS_{n,m}$ is considered as an ideal PS configuration, and the amplitude of each IRS element is set to 1, as follows~\cite{DovelosICC2021}
\begin{align} \label{eq_ps_opt}
    \theta^\star_{n,m} = \frac{2 \pi}{\lambda}(d_{\tx_k,\IRS_{n,m}}+d_{\IRS_{n,m},\rx_l}),
\end{align}
    \begin{align} \label{eq_ps_opt1}
    \kappa^\star_{n,m} = 1, \forall{n,m} .
\end{align}
In the far-field $d_{\tx_k,\IRS_{n,m}}>d_{ray}$, the distance $d_{\tx_k,\IRS_{n,m}}$ using some basic algebra and first order Taylor expansion $(1+a)^x\approx 1+ax$, can be expressed as~\cite{DovelosICC2021}
\begin{align}\label{far_d}
    d_{\tx_k,\IRS_{n,m}}&\approx d_{\tx_k,\IRS_n}-m_xM_x\cos\phi^{\icd}\sin\psi^{\rfl}\nonumber\\&-m_yM_y\sin\phi^{\icd}\sin\psi^{\rfl}.
\end{align}
 For THz communications, the elements of the IRS are densely installed, and we ignore the distance between the IRS elements., which yields
 \begin{align}\label{far_d1}
    d_{\tx_k,\IRS_{n,m}}\approx d_{\tx_k,\IRS_n}.
\end{align}
Similarly $d_{\IRS_{n,m},\rx_l}$ can be written as
 \begin{align}\label{far_d2}
    d_{\IRS_{n,m},\IRS_n}\approx d_{\IRS_n,\rx_l}.
\end{align}
From~\eqref{far_d1} and~\eqref{far_d2}, the PS in~\eqref{eq_ps_opt} is also valid for the far-field communication. Thus, the near-optimal PS configuration at $\IRS_n$ is determined as
\begin{align} \label{eq_opt_Theta}
    \vec{\Theta}^\star_n = \mathrm{diag}[\kappa^\star_{n,1} e^{ j\theta^\star_{n,1}},.., \kappa^\star_{n,m} e^{ j\theta^\star_{n,m}},..,\kappa^\star_{n,M} e^{ j\theta^\star_{n,M}}].
\end{align}
As a result, the T-I-R SINR achieved by the near-optimal PS configuration can be expressed in~\eqref{eq_sinr_e2e2}, at the top of the next page.
\begin{figure*}[!htp]
	\begin{align} \label{eq_sinr_e2e2}
		&\SINR^{\rm ideal}_{\tx_k,\IRS_n,\rx_l}\!\!\!=\frac{\frac{p_k\lambda^2}{16\pi^2}\Big|\sum\limits_{m=1}^M \frac{ \sqrt{ \gain_{\tx_k} \gain_{\rx_l} \gain_{\IRS_{n,m}}(-\vec{r}_{k,n,m})\gain_{\IRS_{n,m}}(\vec{r}_{n,m,l}) e^{-\kappa_{abs}(f)(d_{\tx_k,\IRS_{n,m}}+d_{\IRS_{n,m},\rx_l})}}}{ d_{\tx_k,\IRS_{n,m}} d_{\IRS_{n,m},\rx_l}}
			\Big|^2}{\sum\limits_{\substack{j=1\\ j\neq k}}^K \sum\limits_{\substack{i=1}}^N p_j\big|(\widehat{\vec{h}}_{\tx_j,\IRS_i})^T\vec{\Theta}_i\widehat{\vec{g}}_{\IRS_i,\rx_l}\big|^2\!\!+\!\!\sum\limits_{\substack{j=1}}^K \sum\limits_{\substack{i=1}}^N p_j\big(\sigma_{\widetilde{g}_{\IRS_i,\rx_l}}^2\big|\vec{\widehat{h}}^T_{\tx_j,\IRS_i}\vec{\widehat{h}}_{\tx_j,\IRS_i}\big|\!\!+\!\!\sigma_{\widetilde{h}_{\tx_j,\IRS_i}}^2\big|\vec{\widehat{g}}^T_{\IRS_i,\rx_l}\vec{\widehat{g}}_{\IRS_i,\rx_l}\big|+\sigma_{\widetilde{h}_{\tx_j,\rx_i}}^2\sigma_{\widetilde{g}_{\IRS_i,\rx_l}}^2\big)+\sigma_l^2}.
	\end{align}
	\hrule 
\end{figure*}
\subsection{Optimal IRS Association using Exhaustive Search}\label{es}
{In this subsection, we focus on the IRS pairing problem in~\eqref{eq_opt_prob} to find the optimal allocation matrix $[\vec{\Omega}]_{\tx_k,\IRS_n,\rx_l}$ with the sub-optimal reflection matrix $\vec{\Theta}^\star_n$ obtained from~\eqref{eq_opt_Theta}. In our ES method, the T-I-R sum rate with near-optimal PS is utilized to find the optimal pairing by checking all possible combinations and selecting the optimal one~\cite{fang2016energy,wang2020channel}. Thus, in our model, we have $\frac{N!}{(N-K)!}$ combinations from T to I, and another $L!$ combinations from I to R, yielding $\frac{N!L!}{(N-K)!}$ combinations. The ES method for~\eqref{eq_opt_prob} is shown in Algorithm~\ref{algo_exh}. The allocation matrix and the achievable rate are set to zero at the beginning of the algorithm. Then, we check all the $\frac{N!}{(N-K)!}$ combinations for the T to I links in the first for loop. In addition, for all combinations, we check the $L!$ combinations for the IRS to user links in the second for loop. Furthermore, if the sum rate of the current combination is greater than the existing combination, we select the current combination; otherwise, we keep the existing combination. This process is iterated until we check all combinations, thus reaching the optimal allocation matrix.}
\begin{algorithm}[!tp]
\caption{Exhaustive Search Algorithm for Problem~\eqref{eq_opt_prob}} \label{algo_exh}
 {\bf Input: } $N$, $K$, $R_{\tx_k,\IRS_n,\rx_l}$, combination of set $C(N,K)$ as $\vec{C}_{temp}$, permutation of set $K$ as $\vec{C}_{temp1}$, allocation matrix $[\vec{\Omega}]_{\tx_k,\IRS_n,\rx_l}$\\
{\bf Initialize: } $[\vec{\Omega}]_{\tx_k,\IRS_n,\rx_l} = \emptyset$,  $\sum R_{\tx_k,\IRS_n,\rx_l}=0$ \\
\For {$i \in \frac{N!}{(N-K)!}$}{
 pairing $\tx_k$ with $k^{th}$ element of the $i^{th}$ row in $\vec{C}_{temp}$ as $\s_k \gets \vec{C}_{temp}(i,k)$ \\
 $\vec{C}_{temp}(i,k)$  gives the $\IRS_n$ \\
 \For {$j \in L!$}{
 pairing $\tx_k$ with $k^{th}$ element of the $j^{th}$ row in $\vec{C}_{temp}$ as $\s_k \gets \vec{C}_{temp}(j,k)$ 
 $\vec{C}_{temp}(j,k)$  gives $\rx_l$ \\
 \For {$ l \in K$}{
 $\sum R_{\tx_l,\IRS_n,\rx_l}$ as in equation~\eqref{rateknl}
 }
 {\bf if} {($\sum R_{\tx_k,\IRS_n,\rx_l} \geq \sum R_{\tx_k,\IRS_n,\rx_l}^\star$
 )}\\
 {
 $[\vec{\Omega}]_{\tx_k,\IRS_n,\rx_l}^\star  \gets [\vec{\Omega}]_{\tx_k,\IRS_n,\rx_l} $\\ 
 {\bf else}\\
 $[\vec{\Omega}]_{\tx_k,\IRS_n,\rx_l}^\star  \gets [\vec{\Omega}]_{\tx_k,\IRS_n,\rx_l}^\star $\\ 
 }
 }
}
 {\bf Output: } {optimal allocation matrix $[\vec{\Omega}]_{\tx_k,\IRS_n,\rx_l}^\star$} 
\end{algorithm}
\begin{Remark} \label{lemma2a}
The T-I-R allocation matrix obtained with the ES method as $[\vec{\Omega}]_{\tx_k,\IRS_n,\rx_l}^\star=\underset{\tx_k,\IRS_n,\rx_l}{\arg \max} \{ R_{\tx_k,\IRS_n,\rx_l} \}$ converges to the global optimal solution of problem~\eqref{eq_opt_prob}.
\end{Remark}

Indeed, the optimality of the ES method can be proved by the following conditions. The ES method characteristics dictate that at every iteration, the sum rate for the current allocation matrix is calculated as in step 8 of Algorithm~\ref{algo_exh}. In 9 to 12 of algorithm~\ref{algo_exh}, we check if the sum rate of the current allocation matrix is greater than the previous one and select the one with a higher sum rate. In this way, we get the global optimal allocation matrix $[\vec{\Omega}]_{\tx_k,\IRS_n,\rx_l}^\star$,  which gives the maximum sum rate.
\begin{Theorem} \label{theorem1}
Problem~\eqref{eq_opt_prob} is an NP-hard problem. Furthermore, problem~\eqref{eq_opt_prob}  is NP-hard even if we only consider the T-I-R association with fixed PS at the IRSs. 
\end{Theorem}
\begin{IEEEproof}
Based on the computational complexity theory, to show that problem~\eqref{eq_opt_prob}  is NP-hard, we follow the following two steps: 1) select an appropriate known NP-hard problem, and 2) prove the optimization problem and the known NP-hard problem have the same objective values.
First, 3D matching is known to be NP-hard. Then, considering an instant with fixed PS at the IRSs in~\eqref{eq_opt_prob} , which makes the optimization problem~\eqref{eq_opt_prob} a 3D matching problem. Therefore, the decision problem of the constructed instance is NP-complete, and the corresponding instance is NP-hard. Since a special case (i.e., with fixed PS) of problem~\eqref{eq_opt_prob}  is NP-hard, the original problem in~\eqref{eq_opt_prob}  is also NP-hard.
\end{IEEEproof}
\subsection{The Proposed Low-Complexity Solution}
\begin{figure}[!htp]
 \centering
 \includegraphics[width=0.75\linewidth]{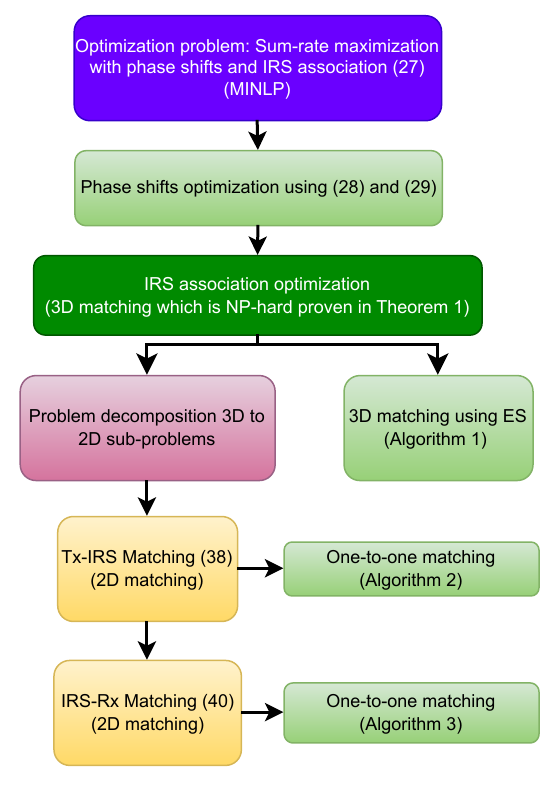}
 \caption{A roadmap for the problem decomposition and the proposed algorithms.}
 \label{roadmap}
 \end{figure}

Every NP-complete problem can be solved by the ES method. However, when the size of the instances grows, the running time becomes forbiddingly large, even for instances of fairly small size.
The complexity of the EC is exponential, and it is non-trivial to obtain an optimal matching in polynomial-time complexity. Therefore, we are motivated to design an efficient solution.
Specifically, we first find the near-optimal IRS reflection coefficients and then decompose the 3D allocation problem into 2D sub-problems and reformulate the allocation optimization problem. The proposed solution is composed of two steps: ($i$) applying the near-optimal PS configuration in \eqref{eq_opt_Theta} for every possible pair and ($ii$) proposing a matching-theory-based low-complexity pairing algorithm. The roadmap for the proposed problem decomposition and the proposed algorithms to the corresponding sub-problems is shown in Fig.~\ref{roadmap}.

Since the IRS is neither a receiver nor a transmitter, $\Xi_{\tx_k,\IRS_n, :}$ represents the pseudo $\SINR$s from $\s_k\to \IRS_n$ estimated based on the imperfect CSI of the $\s_k\to \IRS_n$ link which can be expressed as
	\begin{align} \label{eq_sinr_sr}
		&\Xi_{\tx_k,\IRS_n,:}\triangleq\frac{p_k\big|(\widehat{\vec{h}}_{\tx_k,\IRS_n})^T\widehat{\vec{h}}_{\tx_k,\IRS_n}\big|}{\sum\limits_{\substack{j=1\\ j\neq k}}^K p_j\big|(\widehat{\vec{h}}_{\tx_j,\IRS_n})^T\widehat{\vec{h}}_{\tx_j,\IRS_n}\big|\!+\!\sum\limits_{\substack{j=1}}^K p_j\sigma_{\widetilde{h}_{\tx_j,\IRS_n}}^2\!\!+\!\sigma^2}.
	\end{align}
Here, the symbol ``$:$'' implies that the third dimension of $\vec{\Omega}$ is not taken into account. Furthermore, $\Xi_{{\tx_{k^\star}},{\IRS_{n^\star}},{\rx_l}}$ represents the pseudo $\SINR$ 
 from $\s_{k^\star}\to \IRS_{n^\star}\to \rx_l$ estimated based on the imperfect CSI of the network which can be expressed as 
\begin{align} \label{eq_sinr_ru}
		&\Xi_{\tx_{k^\star},\IRS_{n^\star},\rx_l}\triangleq\frac{p_{k^\star}\big|(\widehat{\vec{h}}_{\tx_{k^\star},\IRS_{n^\star}})^T\vec{\Theta}_{n^\star}\widehat{\vec{g}}_{\IRS_{n^\star},\rx_l}\big|^2}{\sum\limits_{\substack{j=1\\j\neq k}}^K \sum\limits_{\substack{i=1}}^N p_j\big|(\widehat{\vec{h}}_{\tx_j,\IRS_i})^T\vec{\Theta}_i\widehat{\vec{g}}_{\IRS_i,\rx_l}\big|^2+\mathcal{X}+\sigma_l^2},
	\end{align}
 where $\mathcal{X}\!\!=\!\!\sum\limits_{\substack{j=1}}^K \sum\limits_{\substack{i=1}}^N p_j\big(\sigma_{\widetilde{g}_{\IRS_i,\rx_l}}^2\big|\vec{\widehat{h}}^T_{\tx_j,\IRS_i}\vec{\widehat{h}}_{\tx_j,\IRS_i}\big|~+~\sigma_{\widetilde{h}_{\tx_j,\IRS_i}}^2\big|\vec{\widehat{g}}^T_{\IRS_i,\rx_l}\vec{\widehat{g}}_{\IRS_i,\rx_l}\big|+\sigma_{\widetilde{h}_{\tx_j,\IRS_i}}^2\sigma_{\widetilde{g}_{\IRS_i,\rx_l}}^2\big)$.
 We first introduce the following Lemma.
\begin{Lemma} \label{lemma_maxmin}
The pseudo sum rate achieved with allocation matrix obtained by the proposed algorithm for the decomposed sub-problems $\sum\limits_{k^\star,n^\star, k} \Lambda_{\tx_{k^\star},\IRS_{n^\star},\rx_l}([\widehat{\vec{\Omega}}]_{\tx_k,\IRS_n,\rx_l}^\star)$ is a sub-optimal solution to problem~\eqref{eq_opt_prob}, upper bounded by the sum rate achieved with allocation matrix obtained by the ES method $\sum\limits_{\tx_k,\IRS_n,\rx_l}  R_{\tx_k,\IRS_n,\rx_l} ([\vec{\Omega}]_{\tx_k,\IRS_n,\rx_l}^\star)$.
\end{Lemma}
\begin{IEEEproof}
The proof is given in Appendix~\ref{ap_lemma2}
\end{IEEEproof}
Invoking Lemma~\ref{lemma_maxmin}, the original problem can be decomposed into two sub-problems, where the obtained solutions are sub-optimal solution to problem~\eqref{eq_opt_prob}.

\subsubsection{Tx-IRS Matching Problem}

Considering the T-I matching as a bipartite graph matching problem, we take into account the first and second dimensions of $\vec{\Omega}$. From the relationship in \eqref{eq_sinr_proof}, we define the following pseudo quantity to represent the potential sum rate of the incident channels 
\begin{align}\label{ratekn} 
    \Lambda_{\tx_k,\IRS_n , :} ([\vec{\Omega}]_{\tx_k,\IRS_n , :}) \triangleq \log_2(1+ \Xi_{\tx_k,\IRS_n , :}([\vec{\Omega}]_{\tx_k,\IRS_n , :})).
\end{align}
Using \eqref{ratekn}, the first sub-problem of the decomposed problem can be expressed as 
\begin{subequations}\label{sp1}
\begin{alignat}{2}
&\underset{[\vec{\Omega}]_{\tx_k,\IRS_n , :}}{\mathrm{maximize}}
&\quad 
&  \textstyle  \sum_{k=1}^K   \Lambda_{\tx_k,\IRS_n , :} ([\vec{\Omega}]_{\tx_k,\IRS_n , :}) , \label{sp1_optProb}\\
&\text{subject to} 
&& \textstyle   \sum_{k=1}^K [\vec{\Omega}]_{\tx_k,\IRS_n, :} = 1  , \forall{n}, \label{sp1_constraint1}\\
&&&  \textstyle  \sum_{n=1}^N [\vec{\Omega}]_{\tx_k,\IRS_n, :} = 1  , \forall{k}, \label{sp1_constraint2}\\
&&& [\vec{\Omega}]_{\tx_k,\IRS_n, :} = \{0,1\}, \forall k, n, \label{sp1_constraint4}
\end{alignat}
\end{subequations}
where $[\vec{\Omega}]_{\tx_k,\IRS_n, :}$ is the allocation matrix where each element is a 2-dimension, i.e., in this phase, the transmitter to IRS allocation is perform. The constraint~\eqref{sp1_constraint1} ensures that each $\s_k$ is paired to one $\IRS_n$ only. Constraint~\eqref{sp1_constraint2} guarantees that each IRS is allocated to one transmitter. Thus, constraint~\eqref{sp1_constraint1} and \eqref{sp1_constraint2} ensure a one-to-one T-I matching. In this phase we achieve the matching between $\tx_{k^\star}$ and $\IRS_{n^\star}$, and the achieved allocation matrix can be denoted as $[\vec{\Omega}]_{\tx_{k^\star},\IRS_{n^\star} , :}$.
\subsubsection{IRS-Rx Matching Problem}
After pairing  T-I, the allocation matrix $[\vec{\Omega}]_{\tx_{k^\star},\IRS_{n^\star} , :}$ is utilize  and consider as $\tx_{k^\star}$ is paired with $\IRS_{n^\star}$. Consequently, we formulate the I-R pairing problem to maximize the pseudo sum rate in this phase as follows
\begin{align} \label{ratenl}
    &\Lambda_{\tx_{k^\star},\IRS_{n^\star}, \rx_l} ([\vec{\Omega}]_{\tx_{k^\star},\IRS_{n^\star},\rx_l})\nonumber\\&\quad\quad\quad\quad \triangleq \log_2(1+ \Xi_{{\tx_{k^\star}},{\rx_n^\star},{\rx_l}}([\vec{\Omega}]_{\tx_{k^\star},\IRS_{n^\star},\rx_l} )).
\end{align}
Using \eqref{ratenl}, the second sub-problem can be formulated as
\begin{subequations}\label{sp2}
\begin{alignat}{2}
&\underset{[\vec{\Omega}]_{\tx_{k^\star},\IRS_{n^\star},\rx_l}}{\mathrm{maximize}}
&\quad 
& \textstyle  \sum_{l=1}^{L}  \Lambda_{\tx_{k^\star},\IRS_{n^\star},\rx_l}([\vec{\Omega}]_{\tx_{k^\star},\IRS_{n^\star},\rx_l} ), \label{sp2_optProb}\\
&\text{subject to} 
&& \textstyle  \sum_{n^\star=1}^N [\vec{\Omega}]_{\tx_{k^\star},\IRS_{n^\star},\rx_l} = 1  , \forall{k}, \label{sp2_constraint1}\\
&&&  \textstyle  \sum_{l=1}^{L} [\vec{\Omega}]_{\tx_{k^\star},\IRS_{n^\star},\rx_l} = 1  , \forall{n^\star}, \label{sp2_constraint2}\\
&&&  [\vec{\Omega}]_{\tx_{k^\star},\IRS_{n^\star},\rx_l} = \{0,1\}, \forall{n^\star,k}, \label{sp2_constraint4}
\end{alignat}
\end{subequations}
where $[\vec{\Omega}]_{\tx_{k^\star},\IRS_{n^\star},\rx_l}$ is the allocation matrix from I-R link with obtained T-I allocation matrix from first sub-problem. Constraint~\eqref{sp2_constraint1} ensures that each IRS $\IRS_{n^\star}$ is paired to only one Rx $\rx_l$ and~\eqref{sp2_constraint2} guarantees that each IRS is allocated to one Rx. Thus,~\eqref{sp2_constraint1} and \eqref{sp2_constraint2} ensure a one-to-one matching. The implementation of the proposed algorithm is presented in what follows.
\section{Proposed Gale-Shapley Algorithm-based Solution} \label{proposed}
IRS-aided networks require a massive number of reflecting elements $M$, especially at higher frequencies, when the number of elements can reach $100 \times 100$. In this context, our model contains multiple distributed IRSs with a large number of reflecting elements. Thus, the ES method is not an efficient scheme, even for 2D allocation problems. Therefore, to optimize the 2D allocation problems in~\eqref{sp1} and~\eqref{sp2}, we propose a Gale-Shapley algorithm based~\cite{gale1962college} solution, presented in the following sections.
\begin{Definition} \label{def_mat}
A one-to-one matching between two disjoint sets $K$ and $N$ can be represented by a one-to-one correspondence $\mu(.)$, where $k\in K$ is mapped to $n \in N$ if and only if $n$ is also mapped to $k$, i.e., $m=\mu(n)$ and $n=\mu(m)$.
\end{Definition}
\subsection{Phase I: Tx-IRS Matching Problem} \label{sirs}
The proposed scheme for~\eqref{sp1} is elaborated in two stages: configuration of a priority matrix at Tx and IRS for each T-I link, and T-I allocation as a matching problem.
\subsubsection{Tx and IRS Priority Matrix Configuration}
The first sub-problem, i.e.,~\eqref{sp1}, seeks to maximize a pseudo quantity between T-I link through T-I pairing. The pseudo quantity between $\s_k$ and $\IRS_n$ is denoted as $\Lambda_{\tx_k,\IRS_n , :}$, and given as
\begin{align} \label{pr1}
    \Lambda_{\tx_k,\IRS_n , :} \triangleq \log_2(1+ {\Xi}_{\tx_k, \IRS_n, :}).
\end{align}

The transmitter priorities are obtained based on the pseudo quantity $\Lambda_{\tx_k, \IRS_n, :}$ between each IRS and the transmitter. After calculating the pseudo quantity between each transmitter and all the IRSs, and storing them in a matrix $[\vec{\Lambda}]_{\tx_k,\IRS_n,:}^{\s}$, the transmitter priority matrix is constructed. Furthermore, the channel reciprocity holds for the T-I channel~\cite{tang2021channel}, thus the pseudo quantity between each IRS and all the transmitters is the transpose of the matrix $[\vec{\Lambda}]_{\tx_k,\IRS_n, :}^{\s}$, which we denote as $[\vec{\Lambda}]_{\IRS_n,\tx_k, :}^{\IRS}$. The priority matrix at transmitters/IRSs is constructed with the highest pseudo quantity offered by the IRSs/transmitters at the top and the lowest pseudo quantity offered by the one at the bottom, as shown in Algorithm~\ref{algo:1} steps $2$ to $6$. Additionally, the dimension of the transmitter priority matrix $[\vec{\Upsilon}]_{\tx_k,\IRS_n, :}^\s$ is $K \times N$ and the IRS priority matrix $[\vec{\Upsilon}]_{\IRS_n,\tx_k, :}^\IRS$ is $N \times K$. The priority relations for the \textit{\textbf{proposer}} to the \textit{\textbf{responders}} are defined in Definition~\ref{def_pkn}.
\begin{Definition} \label{def_pkn}
The \textit{\textbf{proposer}} $P$ prefers the \textit{\textbf{responder}} $R$ to the \textit{\textbf{responder}} $R^\star$ is denoted as  $R \succ^{{P}} R^\star$, where $R \neq R^\star$, if  $\Lambda_{R,P} >\Lambda_{R^\star,P}$.
\end{Definition}
\subsubsection{Proposed Tx-IRS Allocation Algorithm}\label{proposal1}
\begin{algorithm}[!t]
  \caption{Proposed Tx-IRS Allocation Algorithm for Problem~\eqref{sp1}}
  \label{algo:1}
  \DontPrintSemicolon{
  \KwIn{ $K$, $N$, $\Lambda_{\tx_k,\IRS_n , :}^{\s}$, $\Lambda_{\IRS_n,\tx_k , :}^{\IRS}$, $[\vec{\Upsilon}]_{\tx_k,\IRS_n , :}^{\s}$, $[\vec{\Upsilon}]_{\IRS_n,\tx_k , :}^{\IRS}$, set of unmatched transmitters $\Pi_\s$, $[\vec{\Omega}]_{\tx_k,\IRS_n , :}$}
 {\bf Initialize }$[\vec{\Upsilon}]_{\tx_k,\IRS_n , :}^\s = \emptyset$,
 $[\vec{\Upsilon}]_{\IRS_n,\tx_k , :}^\IRS = \emptyset$, $[\vec{\Omega}]_{\tx_k,\IRS_n , :} = \emptyset$;\;
\underline{\textbf{Priority Matrix Configuration:}}\\
\For {$k\in K$}{
$\Lambda_{\tx_k,\IRS_n , :}, \quad \forall\hspace{2mm} n \in N$ and store in $[\vec{\Lambda}]_{\tx_k,\IRS_n , :}^{\s}$\;
}
$[\vec{\Lambda}]_{\IRS_n,\tx_k , :}^{\IRS} = [[\vec{\Lambda}]_{\tx_k,\IRS_n , :}^{\s}]^T$\;
[$[\vec{\Lambda}]_{\tx_k,\IRS_n, :}^{\s}$,$[\vec{\Upsilon}]_{\tx_k,\IRS_n, :}^\s$] = sort ($[\vec{\Lambda}]_{\tx_k,\IRS_n , :}^{\s}$,2,descend)\;
[$[\vec{\Lambda}]_{\IRS_n,\tx_k  , :}^{\IRS}$,$[\vec{\Upsilon}]_{\IRS_n,\tx_k, :}^\IRS$] = sort ($[\vec{\Lambda}]_{\IRS_n,\tx_k, :}^{\IRS}$,2,descend)\;
\underline{\textbf{Tx-IRS Allocation:}}\\
\While {(either $\Pi_\s$ $\neq \emptyset$ or transmitters not rejected by all IRSs)}{
\For {$\s_{k'}\in \Pi_\s$}{
propose to the highest priority IRS in $[\vec{\Upsilon}]_{\tx_k,\IRS_n , :}^{\s}$\;
element of $[\vec{\Omega}]_{\tx_k,\IRS_n , :} = 1$\;
}
\For {$\IRS_n\in N$}{
{\bf if} {($\IRS_n \notin [\vec{\Omega}]_{\tx_k,\IRS_n , :}$ )}\;
{
$[\vec{\Omega}]_{\tx_k,\IRS_n, :} \gets [\vec{\Omega}]_{\tx_k,\IRS_n, :} \cup (\s_{k'},\IRS_n) $\; 
}
{\bf else if} {($\Lambda_{\IRS_n,\tx_{k'},:}^{\IRS} >  \Lambda_{\IRS_n,\tx_k , :}^{\IRS}$)}\;{
$[\vec{\Omega}]_{\tx_k,\IRS_n, :} \gets [\vec{\Omega}]_{\tx_k,\IRS_n , :} \cup (\s_{k'},\IRS_n)$ \;
{\bf else}\;
$[\vec{\Omega}]_{\tx_k,\IRS_n, :} \gets [\vec{\Omega}]_{\tx_k,\IRS_n , :} \cup (\s_{k},\IRS_n)$ \;
}
}
  }
{\bf Output: }T-I allocation matrix $[\vec{\overline{\Omega}}]_{\tx_{k^\star},\IRS_{n^\star} , :}$
}
\end{algorithm}
We deploy a large number of IRSs in the network to assist the communication and assume that the number of IRSs is greater than the number of Txs (i.e., $N> K$). We explore T-I allocation algorithms as one-to-one matching in the first phase of our proposal. Each transmitter can be allocated to at most one IRS in one-to-one matching. The remaining IRSs, which are not allocated to any transmitter, are considered inactive for this round. The algorithm, shown in Algorithm~\ref{algo:1}, is illustrated as a series of attempts from transmitters to IRSs. The IRS can either be associated with the transmitter or remain vacant during this process. Each transmitter node proposes pairing with the highest priority IRS in their priority list until the transmitter has either been paired or rejected by all IRS. Transmitters rejected by the IRS are not allowed to try again for the same IRS. The IRS is immediately engaged if the transmitter proposes a free IRS; however, if the transmitter proposes an IRS that is already engaged, the IRS compares the new and current transmitter and gets engaged with the best one, i.e., the one with higher pseudo quantity link. Thus, if the IRS favors the current transmitter, the new proposal will be rejected. Alternatively, if the IRS prefers the new proposal, then the IRS will break the engagement with the current transmitter and accept the new one. The process is repeated until all transmitter are engaged or all options are considered.
\begin{Lemma} \label{theorem1a}
The proposed \textit{\textbf{proposer}}-\textit{\textbf{responder}} allocation algorithm terminates in polynomial time, i.e., if there are $K$ \textit{\textbf{proposers}} and $N$ \textit{\textbf{responders}} then the algorithm is terminated after at most $K \times N$ iterations.
\end{Lemma}
\begin{IEEEproof}
In each iteration, each unmatched \textit{\textbf{proposer}} proposes to a \textit{\textbf{responder}} that it has never explored before. Moreover, the \textit{\textbf{responder}} accepts or rejects the new proposal according to the status and preference of the \textit{\textbf{responder}}. As a result, for $N$ \textit{\textbf{responders}} and $K$ \textit{\textbf{proposers}}, we have $K \times N$ possible proposals occuring in the proposed algorithm. This completes the proof of Lemma \ref{theorem1a}.
\end{IEEEproof}
\begin{Lemma} \label{lemma3}
If \textit{\textbf{proposer}} $P$ pairs with \textit{\textbf{responder}} $R$ in the $i^{th}$ iteration, then in every subsequent iteration, \textit{\textbf{responder}} $R$ pairs with a proposer which is at least as good as $P$.
\end{Lemma}

\begin{IEEEproof}
The proof is given in Appendix~\ref{ap_lemma3}.
\end{IEEEproof}
\begin{Theorem} \label{theorem2}
The \textit{\textbf{proposer}}-\textit{\textbf{responder}}  matching achieved for each phase of the proposed algorithm returns a stable matching. Furthermore, the stability is not affected by inverting the proposing order.
\end{Theorem}
\begin{IEEEproof}
The proof is given in Appendix~\ref{ap_theorem2}.  
\end{IEEEproof}
\subsection{Phase II: (Tx, IRS)-Rx Matching Problem} \label{irsu}
The allocation matrix obtained in subsection~\ref{sirs} can be used to achieve the I-R allocation matrix. In this way, we achieve the T-I-R allocation matrix as a sub-optimal solution.
\subsubsection{IRS and Rx Priority Matrix Configuration}
Considering that ($\tx_{k^\star}, \IRS_{n^\star}$) is the pair obtained in the T-I allocation, the data rate offered by $\ue_k$ to the ($\tx_{k^\star}, \IRS_{n^\star}$) channel is given as
\begin{align} \label{pr2}
    \Lambda_{\tx_{k^\star},\IRS_{n^\star},\rx_{k}} \triangleq \log_2(1+ \Xi_{{\tx_{k^\star}},{\IRS_{n^\star}},\rx_l}).
\end{align} 

Combining all the data rates between IRSs and receivers yields a data rate matrix, $[\vec{\Lambda}]_{\tx_{k^\star},\IRS+{n^\star},\rx_l}^\IRS$, and the date rate matrix between each receiver to all the IRSs, $[\vec{\Lambda}]_{\tx_{k^\star},\rx_l,\IRS_{n^\star}}^\rx$, is the transpose of $[\vec{\Lambda}]_{\tx_{k^\star},\IRS_{n^\star},\rx_l}^\IRS$. In this context, the priority matrix is created in descending order of priorities. Accordingly, the Rx/IRS with the highest priority comes first for each IRS/Rx in the priority matrix.  The priority matrix for IRSs and receivers, which we represent as $[\vec{\Upsilon}]_{\tx_{k^\star},\IRS_{n^\star},\rx_l}^\IRS$ and $[\vec{\Upsilon}]_{\tx_{k^\star},\rx_l,\IRS_{n^\star}}^\rx$, can be calculated as shown in Algorithm~\ref{algo:2} steps $2$ to $6$, respectively.
\subsubsection{Proposed Rx-IRS Allocation Algorithm}\label{proposal2}
\begin{algorithm}[!tp]
   \caption{Proposed Rx-IRS Allocation Algorithm for Problem~\eqref{sp2}}
  \label{algo:2}
  \DontPrintSemicolon{
  \KwIn{$L$, $\Lambda_{\tx_{k^\star},\IRS_{n^\star},\rx_l}^{\IRS}$,$\Lambda_{\tx_{k^\star},\rx_{k},\IRS_{n^\star}}^{\rx}$, $[\vec{\Upsilon}]_{\tx_{k^\star},\rx_l,\IRS_{n^\star}}^{\rx}$, $[\vec{\Upsilon}]_{\tx_{k^\star},\IRS_{n^\star},\rx_l}^{\IRS}$, set of unmatched receivers $\Pi_\rx$, $[\vec{\Omega}]_{\tx_{k^\star},\IRS_{n^\star},\rx_l}$}
  
 {\bf Initialize } $[\vec{\Upsilon}]_{\tx_{k^\star},\IRS_{n^\star},\rx_l}^\IRS = \emptyset$,
 $[\vec{\Upsilon}]_{\tx_{k^\star},\rx_l,\IRS_{n^\star}}^\rx = \emptyset$, $[\vec{\Omega}]_{\tx_{k^\star},\IRS_{n^\star},\rx_l} = \emptyset$;\;
\underline{\textbf{Priority Matrix Configuration:}}\\
 \For {$l\in L$}{
$\Lambda_{\tx_{k^\star},\IRS_{n^\star},\rx_l}, \quad \forall\hspace{2mm} k \in K$ and store in $[\vec{\Lambda}]_{\tx_{k^\star},\IRS_{n^\star},\rx_{k}}^{\IRS}$\;
}
$[\vec{\Lambda}]_{\tx_{k^\star},\rx_l,\IRS_{n^\star}}^{\rx}= [[\vec{\Lambda}]_{\tx_{k^\star},\IRS_{n^\star},\rx_l}^{\IRS}]^T$\;
[$[\vec{\Lambda}]_{\tx_{k^\star},\IRS_{n^\star},\rx_l}^{\IRS}$,$[\vec{\Upsilon}]_{\tx_{k^\star},\IRS_{n^\star},\rx_l}^\IRS$] = sort ($[\vec{\Lambda}]_{\tx_{k^\star},\IRS_{n^\star},\rx_l}^{\IRS}$,2,descend)\;
[$[\vec{\Lambda}]_{\tx_{k^\star},\rx_l,\IRS_{n^\star}}^{\rx}$,$[\vec{\Upsilon}]_{\tx_{k^\star},\rx_l,\IRS_{n^\star}}^\rx$] = sort ($[\vec{\Lambda}]_{\tx_{k^\star},\rx_l,\IRS_{n^\star}}^{\rx}$,2,descend)\;
\underline{\textbf{Rx-IRS Allocation:}}\\
\While {(either $\Pi_\ue$ $\neq \emptyset$ or receivers not rejected by all IRSs)}{
\For {$\ue_{k'}\in \Pi_\ue$}{
propose to the highest priority IRS in $[\vec{\Upsilon}]_{\tx_{k^\star},\rx_l,\IRS_{n^\star}}^{\rx}$\;
element of $[\vec{\Omega}]_{\tx_{k^\star},\IRS_{n^\star},\rx_l} = 1$, Remove $\IRS_{n^\star}$ from $[\vec{\Upsilon}]_{\tx_{k^\star},\IRS_{n^\star},\rx_l}^{\IRS}$
}
\For {$\IRS_{n^\star} \in L$}{
{\bf if} {($\IRS_{n^\star} \notin [\vec{\Omega}]_{\rx_{k^\star},\IRS_{n^\star},\rx_l}$ )}\;
{
$[\vec{\Omega}]_{\tx_{k^\star},\IRS_{n^\star},\rx_l} \gets [\vec{\Omega}]_{\tx_{k^\star},\IRS_{n^\star},\rx_l} \cup (\IRS_{n^\star}, \ue_{k'}) $\;
}
{\bf else if} {($\Lambda_{\tx_{k^\star},\IRS_{n^\star},\rx_{k'}}^{\IRS} >  \Lambda_{\tx_{k^\star},\IRS_{n^\star},\rx_{k}}^{\IRS}$)}\;{
$[\vec{\Omega}]_{\tx_{k^\star},\IRS_{n^\star},\rx_l} \gets [\vec{\Omega}]_{\tx_{k^\star},\IRS_{n^\star},\rx_l} \cup (\IRS_{n^\star}, \ue_{k'})$ \;
{\bf else}\;
$[\vec{\Omega}]_{\tx_{k^\star},\IRS_{n^\star},\rx_l} \gets [\vec{\Omega}]_{\tx_{k^\star},\IRS_{n^\star},\rx_l} \cup (\IRS_{n^\star}, \ue_{k})$ \;
}
}
  }
{\bf Output: } T-I-R allocation matrix $[\vec{\overline \Omega}]_{\tx_k,\IRS_n,\rx_l}^\star$
}
\end{algorithm}
After getting the T-I allocation matrix $[\vec{\overline{\Omega}}]_{\tx_{k^\star},\IRS_{n^\star}}$ in the first phase, we propose a matching algorithm, for R-I allocation. We assume that the number of active IRSs and receivers are similar in this phase, and therefore we investigate R-I allocation algorithms as one-to-one matching. The algorithm is illustrated as a series of attempts from receivers to IRSs. The IRS can either be paired with the Rx or remain vacant during this process. A Rx proposes for matching with the highest priority IRS in their priority list until the Rx has been allocated or rejected by each IRS. Receivers rejected by the IRS are not allowed to try again for the same IRS. The IRS is immediately engaged if the Rx proposes a free IRS. However, if the Rx proposes to the already engaged IRS, then the IRS compares the new and current receivers and gets engaged with the one with highest data rate link. This process continues until all receivers are engaged, or all options have been explored.
\subsection{Case Study of The Proposed Solution}
\begin{figure}[!htp]
 \centering
 \includegraphics[width=\linewidth]{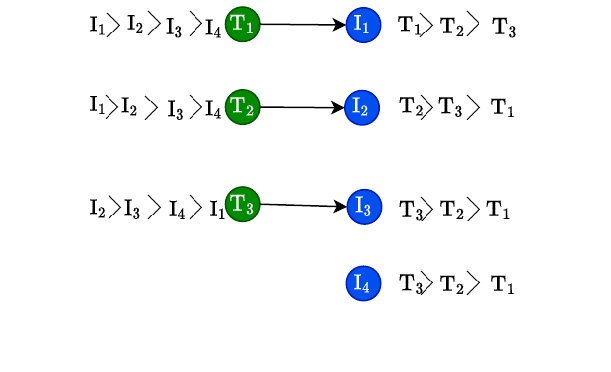}
 \caption{The graph of the Tx - IRS matching problem.}
 \label{usecase1}
 \end{figure}

\begin{table*}
\scriptsize
 \centering
\caption{Priority Matrix of Transmitters, IRSs, and Allocation Rounds of $1^{st}$ Phase from Tx to IRS} \label{Tab:pm}      
\begin{tabular}{cc}
\begin{tabular}{c|c|c|c|c|c|c|c|c|}
\cline{2-9}
             & \multicolumn{2}{c}{\textbf{$1^{st}$ Choice}} & \multicolumn{2}{|c}{\textbf{$2^{nd}$ Choice}}& \multicolumn{2}{|c}{\textbf{$3^{rd}$ Choice}}& \multicolumn{2}{|c|}{\textbf{$4^{th}$ Choice}}\\          
 \cline{2-9} 
                                    & $\IRS$ & {{Eq.\eqref{pr1}}} &
                                    $\IRS$ & {{Eq.\eqref{pr1}}} &
                                    $\IRS$ & {{Eq.\eqref{pr1}}} &$\IRS$ & {{Eq.\eqref{pr1}}} \\ \hline
\multicolumn{1}{|c|}{\textit{\textbf{$\s_1$}} }              & \textit{\textbf{$\IRS_1$}}  & $0.623$ & \textit{\textbf{$\IRS_2$}}  & $0.134$  & \textit{\textbf{$\IRS_3$}} & $0.026$  & \textit{\textbf{$\IRS_4$}} & $0.012$   \\ \hline
\multicolumn{1}{|c|}{\textit{\textbf{$\s_2$}} }              & \textit{\textbf{$\IRS_1$}}  & $0.505$ & \textit{\textbf{$\IRS_2$}}  & $0.448$  & \textit{\textbf{$\IRS_3$}} & $0.044$  & \textit{\textbf{$\IRS_4$}} & $0.022$    \\ \hline
\multicolumn{1}{|c|}{\textit{\textbf{$\s_3$}} }              & \textit{\textbf{$\IRS_2$}}  & $0.203$ & \textit{\textbf{$\IRS_3$}}  & $0.160$  & \textit{\textbf{$\IRS_4$}} & $0.157$  & \textit{\textbf{$\IRS_1$}} & $0.025$    \\  \hline
 
\end{tabular} &
\begin{tabular}{c|c|c|c|c|c|c|}
\cline{2-7}
             & \multicolumn{2}{c}{\textbf{$1^{st}$ Choice}} & \multicolumn{2}{|c}{\textbf{$2^{nd}$ Choice}}& \multicolumn{2}{|c|}{\textbf{$3^{rd}$ Choice}}\\            
 \cline{2-7} 
                                    & $\mathrm{\tx}_k$ & {{Eq.\eqref{pr1}}} &
                                    $\mathrm{\tx}_k$ & {{Eq.\eqref{pr1}}} &
                                    $\mathrm{\tx}_k$ & {{Eq.\eqref{pr1}}} 
                                    \\ \hline
\multicolumn{1}{|c|}{\textit{\textbf{$\IRS_1$}} }              & \textit{\textbf{$\s_1$}}  & $0.623$ & \textit{\textbf{$\s_2$}}  & $0.504$  & \textit{\textbf{$\s_3$}} & $0.025$  
\\ \hline
\multicolumn{1}{|c|}{\textit{\textbf{$\IRS_2$}} }              & \textit{\textbf{$\s_2$}}  & $0.448$ & \textit{\textbf{$\s_3$}}  & $0.203$  & \textit{\textbf{$\s_1$}} & $0.134$    
\\ \hline
\multicolumn{1}{|c|}{\textit{\textbf{$\IRS_3$}} }              
& \textit{\textbf{$\s_3$}}  & $0.160$  & \textit{\textbf{$\s_2$}} & $0.044$  & \textit{\textbf{$\s_1$}} & $0.026$    \\  \hline
\multicolumn{1}{|c|}{\textit{\textbf{$\IRS_4$}} }             
& \textit{\textbf{$\s_3$}}  & $0.157$  & \textit{\textbf{$\s_2$}} & $0.022$  & \textit{\textbf{$\s_1$}} & $0.012$    \\  \hline
\end{tabular}
\cr  (a)  & (b)  
\end{tabular}
\begin{tabular}{c}
\begin{tabular}{c|c|c|c|c|c|c|}
 \cline{2-7} 
                                    & \textit{\textbf{$\IRS_1$}} & \textit{\textbf{$\IRS_2$}} &
                                    \textit{\textbf{$\IRS_3$}} & \textit{\textbf{$\IRS_4$}} &
                 \begin{tabular}[c]{@{}l@{}}$\sum \Lambda_{\tx_k,\IRS_n,:}$\end{tabular} & \textit{\textbf{$\tt LIST$}}  \\ \hline
\multicolumn{1}{|c|}{Round $1$ }              &  \begin{tabular}[c]{@{}l@{}}$\mathtt{CUR} (\s_1)$, $\mathtt{CUR} (\s_2)$, Accept $\s_1$\end{tabular} & \begin{tabular}[c]{@{}l@{}}$\mathtt{CUR} (\s_3)$, Accept $\s_3$\end{tabular} & {$\emptyset$}  & {$\emptyset$} & {$0.826$} & $\{\s_2\}$   \\ \hline
\multicolumn{1}{|c|}{Round $2$ }              &  \begin{tabular}[c]{@{}l@{}}$\mathtt{CUR} (\s_1)$\end{tabular} & \begin{tabular}[c]{@{}l@{}}$\mathtt{CUR} (\s_2)$, Accept $\s_2$, Break $\s_3$\end{tabular} & {$\emptyset$}  & {$\emptyset$} & {$1.071$} & $\{\s_3\}$   \\ \hline
\multicolumn{1}{|c|}{Round $3$ }              &  \begin{tabular}[c]{@{}l@{}}$\mathtt{CUR} (\s_1)$\end{tabular} & \begin{tabular}[c]{@{}l@{}}$\mathtt{CUR} (\s_2)$\end{tabular} &\begin{tabular}[c]{@{}l@{}}$\mathtt{CUR} (\s_3)$, Accept $\s_3$\end{tabular} &{$\emptyset$} & {$1.231$} & {$\emptyset$}   \\ \hline
\end{tabular}
\cr (c)
\end{tabular}
\end{table*}
\begin{table*}
\scriptsize
\centering
\caption{Priority Matrix of Receivers, IRSs, and Allocation Rounds of $2^{nd}$ Phase from Rx to IRS} \label{Tab:2nd}       
\begin{tabular}{cc}
\begin{tabular}{c|c|c|c|c|c|c|}
\cline{2-7}
             & \multicolumn{2}{c}{\textbf{$1^{st}$ Choice}} & \multicolumn{2}{|c}{\textbf{$2^{nd}$ Choice}}& \multicolumn{2}{|c|}{\textbf{$3^{rd}$ Choice}}\\                  
 \cline{2-7} 
                                    & $\IRS$ & {{Eq.\eqref{pr2}}} &
                                    $\IRS$ & {{Eq.\eqref{pr2}}} &
                                    $\IRS$ & {{Eq.\eqref{pr2}}}
                                    \\ \hline
\multicolumn{1}{|c|}{\textit{\textbf{$\ue_1$}} }           
& \textit{\textbf{$\IRS_2$}}  & $0.033$  & \textit{\textbf{$\IRS_1$}} & $0.021$  & \textit{\textbf{$\IRS_3$}} & $0.010$   \\ \hline
\multicolumn{1}{|c|}{\textit{\textbf{$\ue_2$}} }              & \textit{\textbf{$\IRS_1$}}  & $0.040$ & \textit{\textbf{$\IRS_2$}}  & $0.038$  
& \textit{\textbf{$\IRS_3$}} & $0.003$    \\ \hline
\multicolumn{1}{|c|}{\textit{\textbf{$\ue_3$}} }              & \textit{\textbf{$\IRS_1$}}  & $0.058$ & \textit{\textbf{$\IRS_3$}}  & $0.020$  & \textit{\textbf{$\IRS_2$}} & $0.012$  
\\  \hline
\end{tabular} &

\begin{tabular}{c|c|c|c|c|c|c|}
\cline{2-7}
             & \multicolumn{2}{c}{\textbf{$1^{st}$ Choice}} & \multicolumn{2}{|c}{\textbf{$2^{nd}$ Choice}}& \multicolumn{2}{|c|}{\textbf{$3^{rd}$ Choice}}\\         
 \cline{2-7} 
                                    & $\mathrm{\rx}_k$ & {{Eq.\eqref{pr2}}} &
                                    $\mathrm{\rx}_k$ & {{Eq.\eqref{pr2}}} &
                                    $\mathrm{\rx}_k$ & {{Eq.\eqref{pr2}}} 
                                    \\ \hline
\multicolumn{1}{|c|}{\textit{\textbf{$\IRS_1$}} }              
& \textit{\textbf{$\ue_3$}}  & $0.058$  & \textit{\textbf{$\ue_2$}} & $0.040$  & \textit{\textbf{$\ue_1$}} & $0.020$   \\ \hline
\multicolumn{1}{|c|}{\textit{\textbf{$\IRS_2$}} }              & \textit{\textbf{$\ue_2$}}  & $0.038$ & \textit{\textbf{$\ue_1$}}  & $0.033$   
& \textit{\textbf{$\ue_3$}} & $0.012$    \\ \hline
\multicolumn{1}{|c|}{\textit{\textbf{$\IRS_3$}} }              & \textit{\textbf{$\ue_3$}}  & $0.020$ & \textit{\textbf{$\ue_1$}}  & $0.010$ 
& \textit{\textbf{$\ue_2$}} & $0.003$    \\  \hline
\end{tabular}
\cr  (a)  & (b)
\end{tabular}
\begin{tabular}{c}
\begin{tabular}{c|c|c|c|c|c|}
 \cline{2-6} 
                                    & \textit{\textbf{$\IRS_1$}} & \textit{\textbf{$\IRS_2$}} &
                                    \textit{\textbf{$\IRS_3$}} & 
                 \begin{tabular}[c]{@{}l@{}}$\sum\Lambda_{\tx_{k^\star},\IRS_{n^\star},\rx_l}$ \end{tabular} & \textit{\textbf{$\tt LIST$}}  \\ \hline
\multicolumn{1}{|c|}{Round $1$}              &  \begin{tabular}[c]{@{}l@{}}$\mathtt{CUR} (\ue_2)$, $\mathtt{CUR} (\ue_3)$, Accept $\ue_3$\end{tabular} & \begin{tabular}[c]{@{}l@{}}$\mathtt{CUR} (\ue_2)$, $\mathtt{CUR} (\ue_1)$, Accept $\ue_1$\end{tabular} & $\emptyset$ & {$0.091$} & $\{\ue_2\}$   \\ \hline
\multicolumn{1}{|c|}{Round $2$}              &  \begin{tabular}[c]{@{}l@{}}$\mathtt{CUR} (\ue_3)$\end{tabular} & \begin{tabular}[c]{@{}l@{}}$\mathtt{CUR} (\ue_1)$, $\mathtt{CUR} (\ue_2)$, Accept $\ue_2$, Break $\ue_1$ \end{tabular} & $\emptyset$ & $0.096$ & $\{\ue_1\}$ 
 \\ \hline
\multicolumn{1}{|c|}{Round $3$}              &  \begin{tabular}[c]{@{}l@{}}$\mathtt{CUR} (\ue_1)$, $\mathtt{CUR} (\ue_3)$, Reject $\ue_1$\end{tabular} & \begin{tabular}[c]{@{}l@{}}$\mathtt{CUR} (\ue_2)$\end{tabular} & $\emptyset$ & {$0.096$} & $\{\ue_1\}$   \\ \hline
\multicolumn{1}{|c|}{Round $4$}              &  \begin{tabular}[c]{@{}l@{}}$\mathtt{CUR} (\ue_3)$\end{tabular} & \begin{tabular}[c]{@{}l@{}}$\mathtt{CUR} (\ue_2)$ \end{tabular} & \begin{tabular}[c]{@{}l@{}}$\mathtt{CUR} (\ue_1)$, Accept $\ue_1$\end{tabular}& $0.106$ & $\emptyset$   \\\hline
\end{tabular} 
\cr (c)
\end{tabular}
\end{table*}
In the first phase, T-I allocation is considered for $K=3$ and $N=4$. The priority matrices between the Tx and the IRS and IRS and Tx are tabulated in Tables~\ref{Tab:pm}(a) and Table~\ref{Tab:pm}(b), respectively. The transmitters attempt the IRS based on their priorities. In the first round, transmitters target the $1^{st}$ choice IRS as, $\s_1$, $\s_2$, and $\s_3$ try for $\IRS_1$, $\IRS_1$,  and $\IRS_2$, respectively. At this stage, all the IRSs are free, so if an IRS gets one proposal, it accept it and if it gets more than one proposal, it checks the priority of the proposing transmitters and accepts the one with the highest priority. Here, $\IRS_2$ receive one proposal, thus $\IRS_2$ accept the proposal i.e., $\s_3$. However, $\s_1$ and $\s_2$ both want to pair with $\IRS_1$, which checks its priority list before accepting the proposals. In this context, since the priority of $\s_1$ is higher than that of $\s_2$, $\IRS_1$ accept $\s_1$ and reject $\s_2$. Therefore, in the second round, $\s_2$ proposes to the next IRS on its priority list, which is $\IRS_2$. Since $\IRS_2$ is already paired, $\IRS_2$ checks its preference between existing and new proposal. Here, the priority of $\s_2$ is high, and so $\IRS_2$ break the existing pair and accept with $\s_2$. In the third round, $\s_3$ proposes the next IRS available for pairing in its priority list, thus $\IRS_3$ is allocate to $\s_3$. This process is shown in Table~\ref{Tab:pm}(c), where $\mathtt{CUR}(\s)$ is the highest-ranked IRS in the transmitter list that has not been rejected yet and the list of unmatched transmitters are placed in $\mathtt{LIST}$. After allocating all transmitters, we consider the remaining IRSs in the $\mathtt{LIST}$, which are not allocated as inactive IRSs for the second sub-problem. The graphic view of the T-I matched is shown in Fig.~\ref{usecase1}.

\begin{figure}[!htp]
 \centering
 \includegraphics[width=\linewidth]{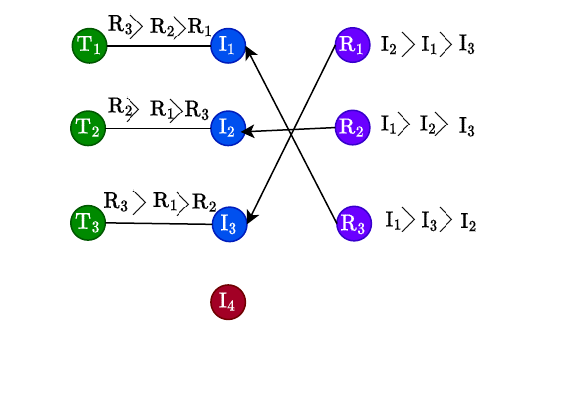}
 \caption{The graph of the (Tx, IRS) - Rx matching problem.}
 \label{usecase2}
 \end{figure}

To demonstrate the second sub-problem after getting the first phase results, we consider an imaginary set of priorities for receivers and IRSs. The priorities of receivers and IRSs' are tabulated in Tables~\ref{Tab:2nd}(a) and~\ref{Tab:2nd}(b), respectively. In the first round, receivers target the $1^{st}$ choice IRS, where $\ue_1$ try for $\IRS_2$, while $\ue_2$, and $\ue_3$ explore $\IRS_1$. Here, $\IRS_2$ being free and targeted by one $\ue$ so it accept the proposal, while $\IRS_1$ checks its priority list before accepting the proposals. Thus, for $\IRS_1$, $\ue_3$'s proposal is accepted since its priority is higher than that of $\ue_2$. In the second round, $\ue_2$ chooses the next IRS in their priority list, i.e., $\IRS_2$. Since the targeted IRS is allocated to $\ue_1$, $\IRS_2$ checks its priority and accept the new proposal and break the existing allocation since its priority is higher than the existing one. In the third round, $\ue_1$ target $\IRS_1$, which is engaged with $\ue_3$. Thus, $\IRS_1$ checks its priority and reject the new proposal since its existing allocated partner priority is higher than the new proposal. Furthermore, in the fourth round, $\ue_1$ chooses $\IRS_3$, and the proposal is accepted since $\IRS_3$ is vacant. This process is shown in Table~\ref{Tab:2nd}(c). After solving both sub-problems the result to the T-I-R allocation is mapped as ($\tx_1\to \IRS_1 \to \rx_3$), ($\tx_2 \to \IRS_2 \to \rx_2$), and ($\tx_3 \to \IRS_3 \to \rx_1$). The graphic view of the T-I-R  matched is shown in Fig.~\ref{usecase2}.
\subsection{Convergence Analysis} \label{convergence}
The convergence of the matching algorithm can be proved following the approach in~\cite{shamaei2018interference}. As stated in the proposed algorithm, proposers (Txs~/~Rxs) that have not been matched at each iteration apply to the next preferred responders (IRSs) on their priority lists. Hence, there exists competition among proposers applying for a responder. The assignment rules in the proposed algorithm indicate that the responder accepts the proposer that maximizes the sum rate. Since proposers only apply to the responder on their priority list once, the request and reject procedure ends when every proposer has matched or has been rejected by all responders. Thus it can be concluded that the proposed allocation algorithm terminates after a few iterations.  
 \begin{figure}[!htp]
 \centering
 \includegraphics[width=\linewidth]{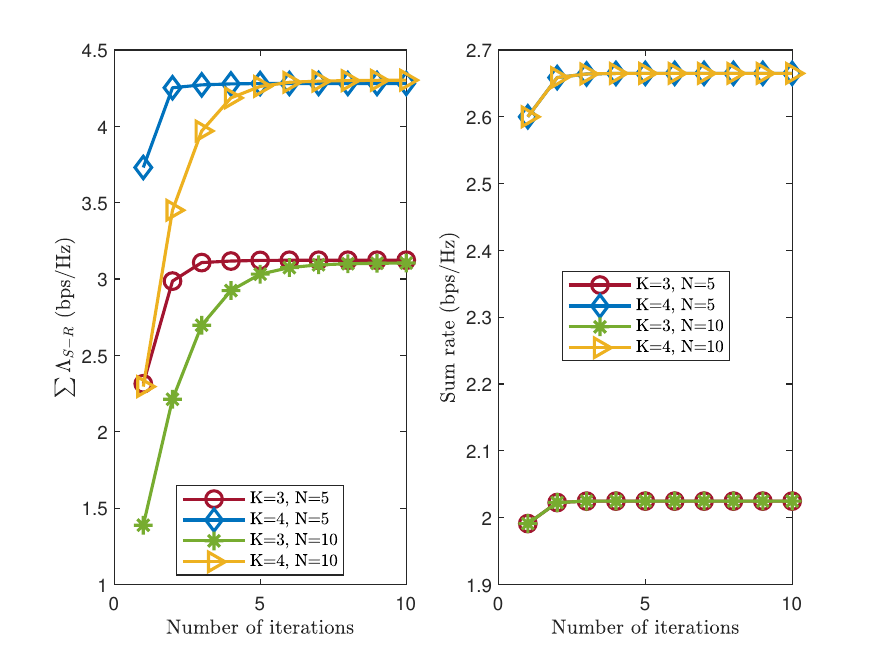}
 \caption{Sum rate versus number of iterations for different number of Tx / Rx pairs with a network area of $20\times20$ $\mathrm{m}^2$, $p_{k} =25$ dBm, and $M=100\times100$.}
 \label{sr_iter}
 \end{figure}
 Fig.~\ref{sr_iter} shows the sum rate for varying numbers of iterations with a different number of Tx / Rx pairs. It can be observed that the convergence rate decreases when the number of Tx / Rx increases since more Tx / Rx pairs need to be paired. Furthermore, when the number of IRSs increases, the convergence rate decreases in the first phase. However, the convergence rate in the second phase remains the same when the number of IRSs increases. This is because we consider that the IRSs allocated in the first phase will participate in the second phase while the rest remain inactive. The convergence of~\eqref{sp1} is shown in the first subplot, and the second subplot demonstrates the convergence of~\eqref{sp2}. Moreover, we observe that the overall algorithm can converge very fast.
\subsection{Computational Complexity Analysis} \label{complexity}
In order to reach the optimal T-I-R pairing, all possible combinations must be searched and paired with the one that maximizes the sum rate in~\eqref{eq_opt_prob}, which is known as ES. The computational complexity of the ES method is $\mathcal{O}(\frac{N!L!}{(N-K)!})$. Meanwhile, the computational complexity of the PES scheme is $\mathcal{O}(\frac{N!}{(N-K)!}+L!)$ and that of the proposed matching-based scheme is $\mathcal{O}((K\times N)+(L\times L))$. Additionally, to compare the computational complexity of different methods, we take the natural logarithm of the linear computational complexity. The logarithm complexity of the ES and PES schemes is $\mathcal{O}(ln(\frac{N!L!}{(N-K)!}))$ and $\mathcal{O}(ln(\frac{N!}{(N-K)!}+L!))$, respectively, which can be simplified using logarithmic properties and Stirling's formula, $\ln (n!)= n\ln n$. Accordingly, the logarithm complexity of the ES is $\mathcal{O}(N\ln{(N)+L\ln(L)-(N-K)\ln(N-K))}$. Moreover, the complexity of the proposed scheme can be expressed in natural logarithm as $\mathcal{O}(\ln{K}+\ln{N}+\ln{L})$, which is much less than that of the ES schemes. 
\section{Results and Discussion} \label{pref}
In this section, we evaluate the performance of the proposed matching-based allocation in terms of sum rate. We implemented the proposed matching-based allocation scheme using MATLAB. In the simulated scenario, explained in Section~\ref{model}, the transmitters, receivers, and IRSs are deployed within the network area. The carrier frequency and the channel BW are set to be $f_c =300$ GHz and $10$ GHz~\cite{zhang2020energy}, respectively. The length and width of each IRS element is $M_x=M_y= 0.4 \lambda$~\cite{najafi2020physics}, where $\lambda$ is the wavelength. The transmit power ($p_\tx$), noise power density ($N_0$), and noise figure (NF) are $25$ dBm~\cite{zhang2020energy}, $-174$ dBm/Hz~\cite{zhang2020energy}, and $10$ dB~\cite{zhang2020energy}, respectively. The AWGN noise power at each receiver is set to be $\sigma_l^2=-N_0+10\log_{10}(BW)+NF=-64$~dBm. Simulation parameters, detailed in Table~\ref{tab:sim} are configured based on existing works. We demonstrate how the transmission power of the transmitter influences the sum rate, the number of IRS elements, the size of the network, and the number of transmitters/receivers. Furthermore, the proposed scheme is compared to the following schemes:
\begin{itemize}
    \item \textbf{Exhaustive Search (ES):} In the ES method, the T-I-R sum rate is utilized to find the optimal pairing by checking all possible combinations and selecting the optimal one, as in Section~\ref{es}.
    \item \textbf{Partial Exhaustive Search (PES):} In the PES scheme, all the possible T-I pairing combinations are explored. Following this allocation, PES explores the optimal I-R matching, which leads to sub-optimal T-I-R matching.
    \item \textbf{Greedy Search (GS):} In the GS method~\cite{xu2015user,sun2021sum}, each T / R selects the IRS with a higher data rate. IRSs receiving multiple proposals are randomly allocated to one of the transmitters/receivers.
    \item \textbf{Nearest Association (NA):} In the NA rule, each transmitter/receiver is associated with the IRS which is closest to it among all IRSs in $N$~\cite{mei2021performance,sun2023irs}.
    \item \textbf{Random Allocation (RA):} The RA scheme~\cite{cui2017optimal} randomly selects the T-I-R allocation matrix.
    \item \textbf{Partial Random Allocation (PRA):}  The PRA scheme randomly selects the T-I allocation matrix, then also randomly selects the I-R allocation matrix.
\end{itemize}
\begin{table}[!htp]
\centering
\renewcommand{\arraystretch}{1}
\caption{Simulation Parameters}
\label{tab:sim}
\begin{tabular}{l l }
\hline

\textbf{Parameters}                                  & \textbf{Values} \\ \hline
Carrier frequency, $f_c$ {[}GHz{]}                   & $300$            
\\Side length of IRS elements & $0.4 \lambda$~\cite{najafi2020physics} \\ 
Absorption coefficient, $\kappa_{abs}(f)$ ${[}m^{-1}{]}$   & $0.0033$ ~\cite{DovelosICC2021}               
\\ 
Transmit power, $p_\s$ {[}dBm{]}                    & $25$~\cite{zhang2020energy}           
\\
Noise power density, $N_0$ {[}dBm/Hz{]} & $-174$~\cite{zhang2020energy}           
          \\ 
Channel bandwidth, $BW$ {[}GHz{]}                     & $10$~\cite{zhang2020energy}           
\\ 
Noise figure, NF {[}dB{]}                           & $10$~\cite{10317892}          
\\ 
Number of IRS elements, $M$                              & $100\times100$~\cite{10317892} \\          
$\tx_k$ coordinate $[m]$&  $x_\tx^k \sim \mathcal{U}(0\,\,20)$, $y_\tx^k \sim \mathcal{U}(0\,\,20)$
\\
$\IRS_n$ coordinate $[m]$&  $x_\IRS^n \sim \mathcal{U}(0\,\,20)$, $y_\IRS^n \sim \mathcal{U}(0\,\,20)$,\\ &$z_\IRS^n \sim \mathcal{U}(0\,\,5)$\\
$\rx_l$ coordinate $[m]$&  $x_\rx^l \sim \mathcal{U}(0\,\,20)$, $y_\rx^l \sim \mathcal{U}(0\,\,20)$\\
Avg. height of transmitters $[m]$ & $1$\\
Avg. height of receivers $[m]$ & $1$
\\ \hline
\end{tabular}
\end{table}
\subsection{Impact of the Transmission Power}
\begin{figure}[!htp]
 \centering
 \includegraphics[width=\linewidth]{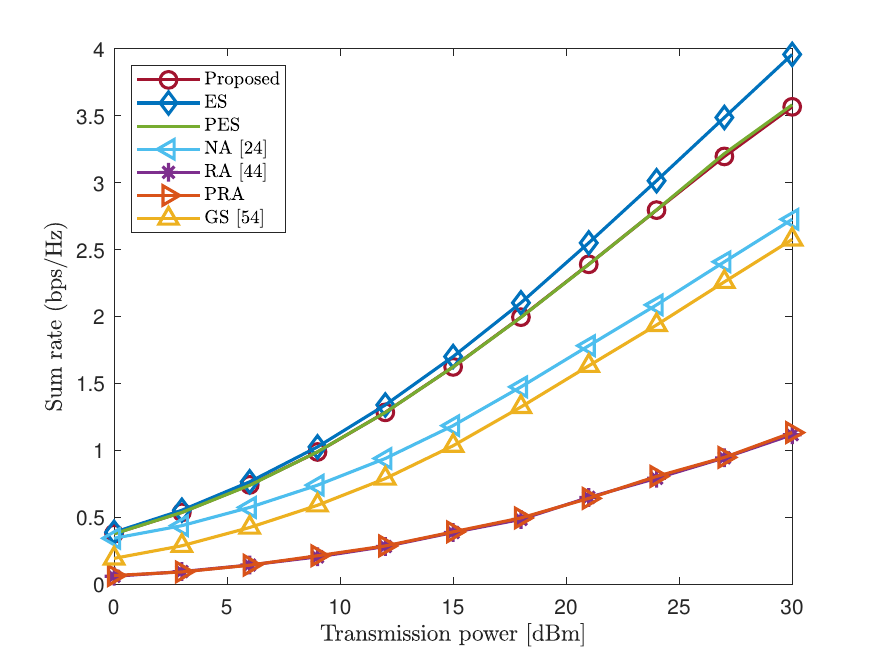}
 \caption{{Sum rate versus transmit power for a network area of $20\times20$ $\mathrm{m}^2$ and $M=100\times100$.}}
 \label{sr_p}
 \end{figure}

We consider $N=5$, $K=3$, and $M=100\times 100$ to illustrate the impact of the transmission power on the performance matrix. Here, the sum rate increases as the transmission power increases for all the cases, as shown in Fig.~\ref{sr_p}. Moreover, the sum rate of the proposed scheme is similar to that of the PES scheme and comparable to that of the ES scheme. In addition, the sum rate of the proposed algorithm outperforms that of the NA, GS, PRA, and RA schemes.
\subsection{Impact of the Number of IRS Elements}
\begin{figure}[!htp]
 \centering
 \includegraphics[width=\linewidth]{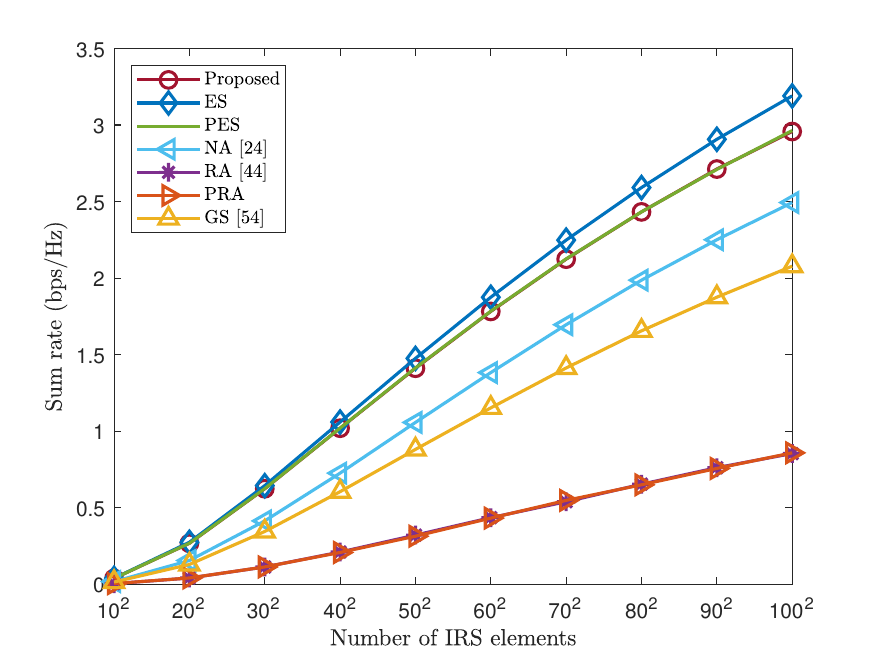}
 \caption{{Sum rate with varying number of IRS elements for a network area of $20\times20$ $\mathrm{m}^2$ and $p_{k} =25$ dBm.}}
 \label{sr_m}
 \end{figure}

To demonstrate the impact of number of reflecting elements on the performance matrix, we consider $N=5$, $K=3$, and $p_k=25 dBm$. The sum rate increases as $M$ increases, as shown in Fig.~\ref{sr_m}. The reason behind this is that when $M$ increases, the total received signal strength also increase, which ultimately improves the sum rate. Moreover, the sum-rate of the proposed algorithm is similar to that of the PES scheme and comparable to the ES method. Furthermore, the proposed scheme achieves a higher sum rate than the NA, GS, RA, and PRA, and this is due to the proper allocation of IRSs to transmitters and receivers.
\subsection{Impact of the IRS Reflecting Efficiency}
\begin{figure}[!htp]
 \centering
 \includegraphics[width=\linewidth]{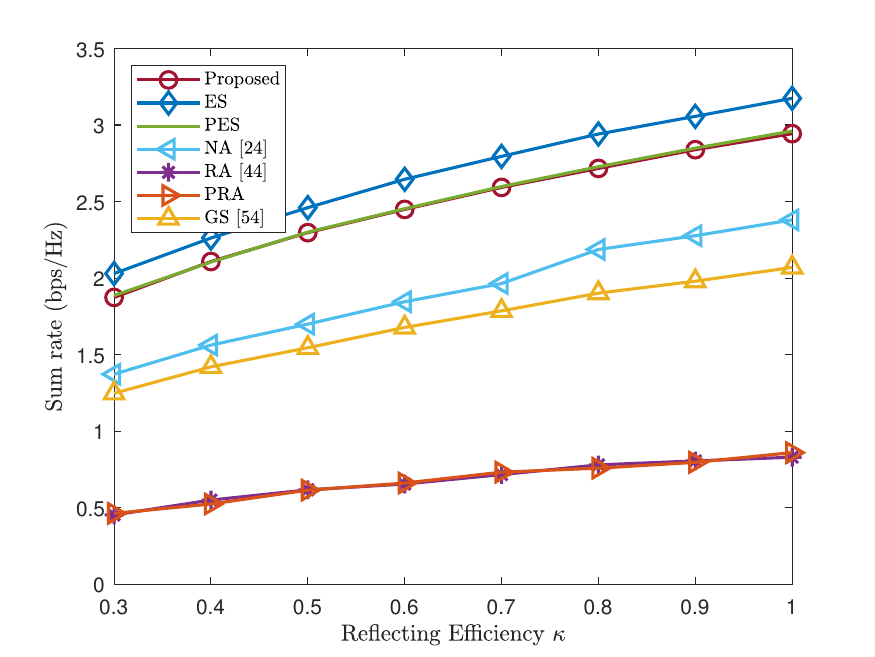}
 \caption{{Sum rate versus  IRS reflecting efficiency for a network area of $20\times20$ $\mathrm{m}^2$ and $p_{k} =25$ dBm.}}
 \label{sr_eff}
 \end{figure}

Fig.~\ref{sr_eff} demonstrates the sum rate in relation to the reflecting efficiency of IRSs.
It can be observed that the reflecting efficiency of the IRS has a substantial impact on the sum rate, as expected. As $\kappa$ increases the sum rate achieved by all schemes increases significantly. This is due to the less power loss caused by signal absorption at IRSs for higher $\kappa$.
\subsection{Impact of the Network Area}
\begin{figure}[!htp]
 \centering
 \includegraphics[width=\linewidth]{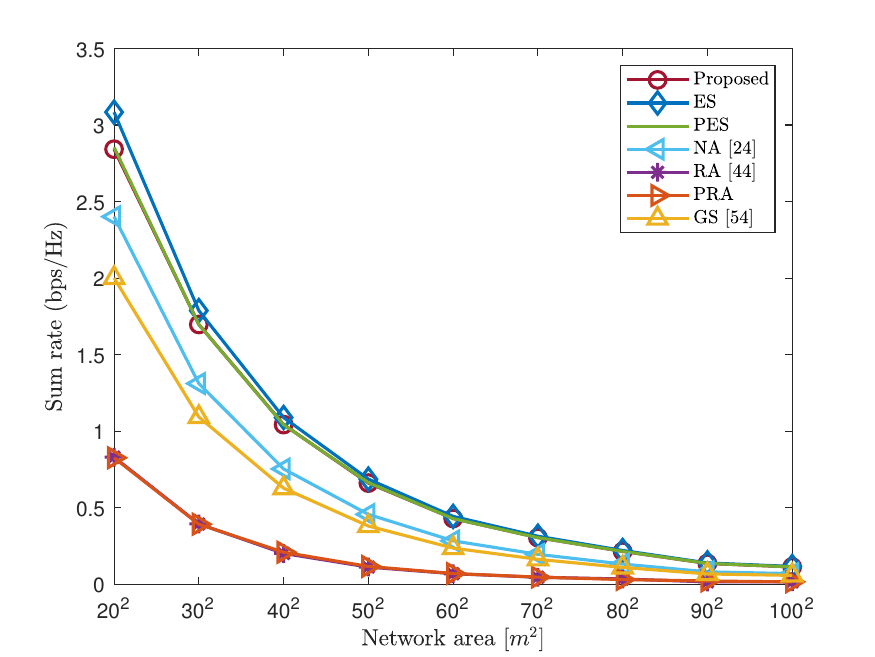}
 \caption{{Sum rate with a varying network area for $p_{k} =25$ dBm and $M~=~100~\times~100$.}}
 \label{sr_a}
 \end{figure}

As shown in~Fig.~\ref{sr_a}, for all schemes, the sum rate decreases as the network size increases due to the increase in the distance between the transmitters and IRSs as well as between the IRSs and receivers. Specifically, the PL is distance-dependent, and therefore the sum rate decreases against increases in the distance. Additionally, for the given network size, the sum rate achieved by the proposed scheme is identical to that of the PES scheme and superior to the sum rate of the NA, GS, PRA, and RA schemes, as depicted in Fig.~\ref{sr_a}.
 \subsection{Impact of Number of Tx/Rx and Frequency}
\begin{figure}[!htp]
 \centering
 \includegraphics[width=\linewidth]{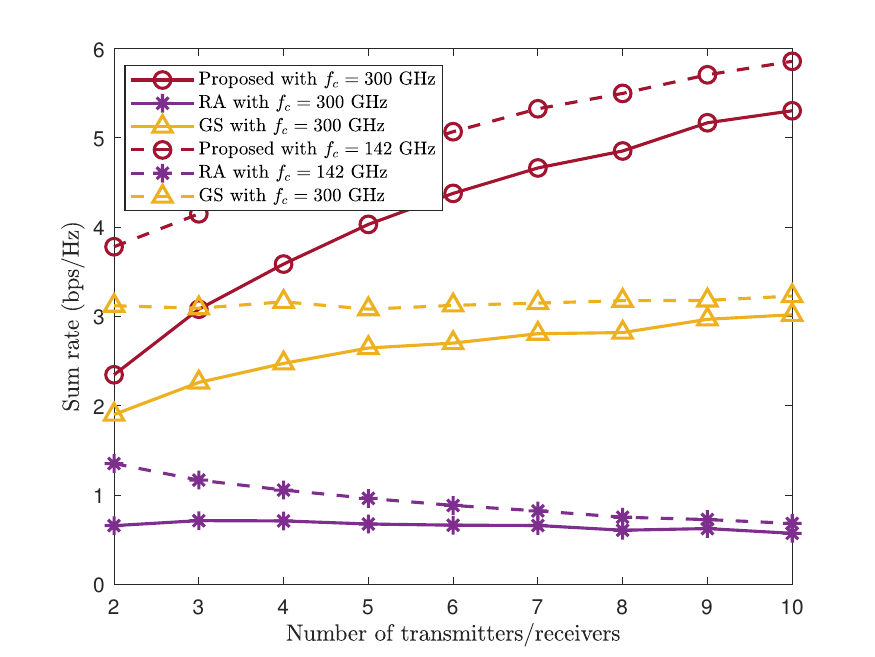}
 \caption{Sum rate with varying number of transmitters/receivers for a network area of $20\times20$ $\mathrm{m}^2$, $p_{k} =25$ dBm, N=10, and $M=100\times100$.}
 \label{sr_u}
 \end{figure}

Fig.~\ref{sr_u} shows the impact of the number of transmitters/receivers on the sum-rate performance for $142$ GHz and $300$ GHz frequencies with a BW of $1$ GHz and $10$ GHz, respectively. 
  The performance of the proposed scheme is superior for both frequencies, which is due to the proper scheduling. In the meantime, the sum rate for the GS scheme remains approximately steady due to the optimal targets by proposers and the random selection of responders. However, the sum rate for the RA scheme decreases due to the increasing interference. Moreover, a higher sum-rate is achieved at $142$ GHz due to lower PL.
  \begin{figure}[!htp]
 \centering
 \includegraphics[width=\linewidth]{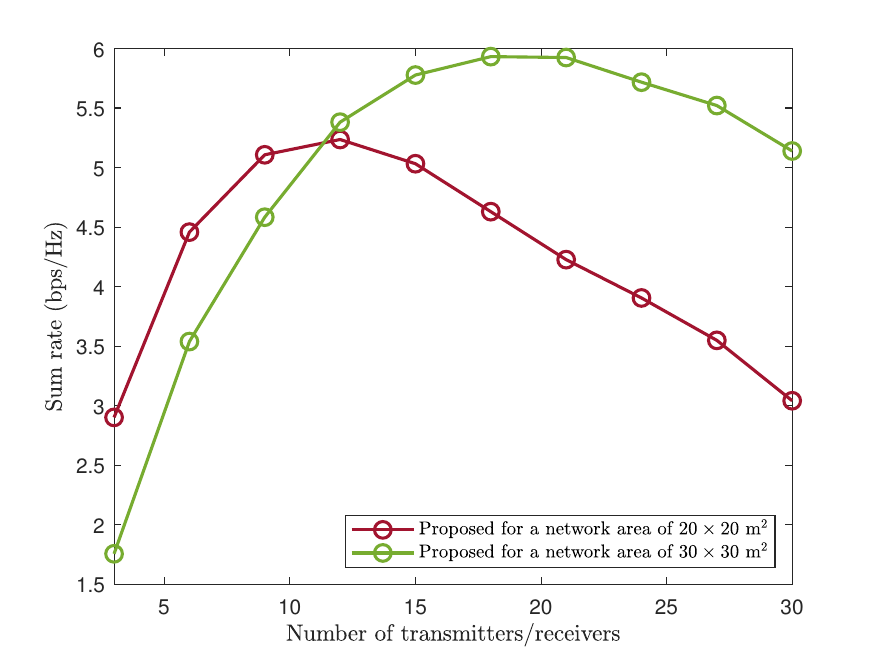}
 \caption{{Sum rate with the varying number of transmitters/receivers.}}
 \label{sr_u_a}
 \end{figure}
 
Fig.~\ref{sr_u_a} demonstrate the impact of the number of transmitters/receivers on the sum rate for two different network area. For the network area of $20\times 20\, m^2$, the peak sum rate is achieved approximately with 10 transmitters/receivers. However, the peak sum rate is achieved approximately with 21 transmitters/receivers for the network area of $30\times 30 \,m^2$. Thus, we conclude that the sum rate depends on the network area and the number of transmitters/receivers.

\begin{figure}[!htp]
 \centering
 \includegraphics[width=\linewidth]{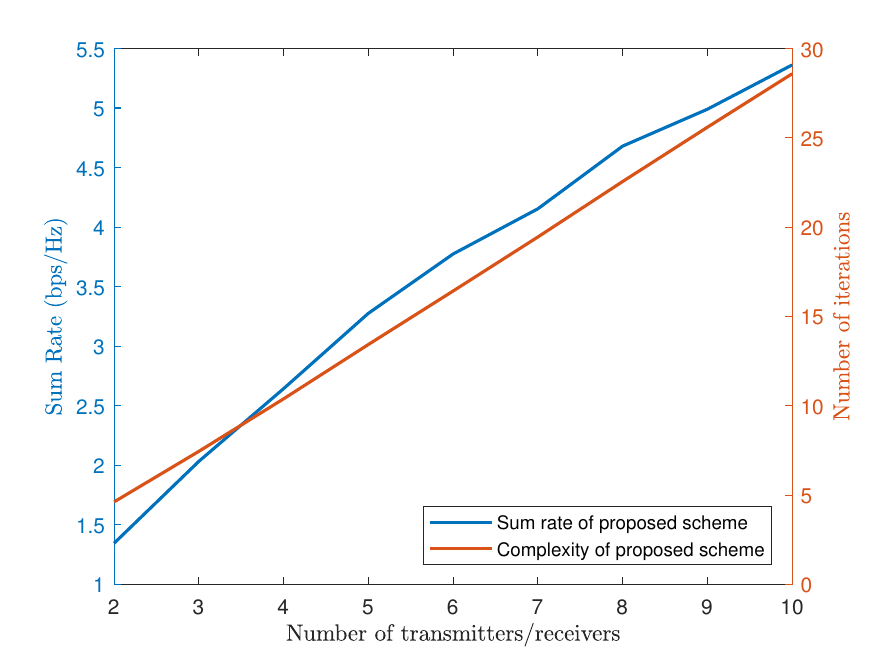}
 \caption{A tradeoff between sum rate and complexity for different numbers of transmitters/receivers with a network area of $20\times20$ $\mathrm{m}^2$, and $M=100\times100$.}
 \label{sr_complexity}
 \end{figure}

 The tradeoff between the sum rate and the computational complexity with a varying number of transmitters/receivers is elaborated in Fig.~\ref{sr_complexity}. It can be observed that the sum rate and the computational complexity increase with increases in the number of Tx / Rx pairs. It can be spotted from Fig.~\ref{sr_complexity} that for $10$ Tx / Rx pairs, the algorithm converges within $30$ iterations and requires less iterations for less than $10$ Tx / Rx.
\section{Conclusions}\label{conc}
In this paper, we considered IRS-aided THz networks with imperfect CSI, where several IRSs are deployed to assist the communication. First, we investigated the PL model for the T-I channel and I-R channel and proceeded to derive the PL model for the T-I-R cascaded channel. We then formulated the T-I-R matching problem to maximize the sum rate, an NP-hard problem. Thus, to solve the NP-hard T-I-R matching problem, we decomposed the problem into two sub-problems, which we solved with our proposed matching-based algorithm. Furthermore, we divided our proposal into two phases. In the first phase, we solved the first T-I matching sub-problem. Then, in the second phase, we achieved I-R matching using the result of the first phase. Simulation results consolidated the potential of the proposed approach for IRS-aided THz communications and demonstrated its superior performance compared to the NA, GS, PRA, and RA schemes. Moreover, provided complexity and convergence analysis verified that the computational complexity of our algorithm is lower than the ES and PES schemes. Possible future research directions based on the results obtained in this paper include the study of the mobility support model, the analysis of near-field communications, and the study of a practical amplitude and phase shift model of IRS reflecting elements.

\appendices
\section{Proof of Lemma~\ref{lemma_gain}} \label{ap_lemma1}
Considering that the incident wave from $\tx_k$ is linearly polarized along the $x$-axis, 
its electric field (E-field) is presented as ${\vec{E}^{\icd}_{k} \!=\! E_{k}^\icd e^{-j \kappa (y\sin \theta^{\icd}_{k} - z\cos \theta^{\icd}_{k})} \vec{e}_{x}}$ \cite{DovelosICC2021}, where $E_{k}^\icd$ [V/m] denotes the amplitude of the incident E-field and $\theta^{\icd}_{k}$ [rad] denotes the incidence angle. We rewrite the E-field as $\vec{E}^{\icd}_{k} = (E^{\icd}_{x, k}, E^{\icd}_{y, k}, E^{\icd}_{z, k}) $. Since ${\vec{e}_{x} = (1, 0,  0)}$, we obtain
\begin{align}
E^{\icd}_{x, k} = E_{k}^\icd
    e^{-j \kappa (y\sin \theta^{\icd}_{k} - z\cos \theta^{\icd}_{k})},
~E^{\icd}_{y, k} = E^{\icd}_{z, k} = 0.
\end{align}

Using Faraday equation, i.e., ${\vec{\nabla} \times \vec{E}^{\icd}_{k} = -j \omega \vec{B}^{\icd}_{k}}$ with ``$\times$'' being the cross product operator, we can determine the incident magnetic B-field ${\vec{B}^{\icd}_{k} \triangleq (B^{\icd}_{x, k}, B^{\icd}_{y, k}, B^{\icd}_{z, k})}$. 
    Here, the left-hand side of the Faraday equation is obtained as
\begin{align}
\vec{\nabla} \times \vec{E}^{\icd}_{k}
= \left(
    0, \frac{\partial E^{\icd}_{x, k}}{\partial z}, - \frac{\partial E^{\icd}_{x, k}}{\partial y}
\right),
\end{align}
where ${\vec{e}_{y} = (0, 1, 0)}$ and ${\vec{e}_{z} = (0, 0, 1)}$.
The partial derivative of $E^{\icd}_{x, k}$ with respect to $z$ and $y$ are obtained as ${j \kappa \cos\theta^{\icd}_k E^{\icd}_{x, k}}$ and ${-j \kappa \sin\theta^{\icd}_k E^{\icd}_{x, k}}$, respectively. 
    Hence, comparing the right-hand side of the Faraday equation, we obtain ${B^{\icd}_{x, k} = 0}$, ${B^{\icd}_{y, k} = - \frac{\kappa}{\omega} \cos \theta^{\icd}_k E^{\icd}_{x, k}}$, and ${B^{\icd}_{z, k} = -\frac{\kappa}{\omega} \sin \theta^{\icd}_k E^{\icd}_{x, k}}$. 
Hence, the magnetic B-field components of the incident wave are obtained~as
\begin{align}
\vec{B}^{\icd}_{k}
    = - \frac{E^{\icd}_k}{\eta} \Big[ \vec{e}_{y} \cos\theta^{\icd}_{k} + \vec{e}_{z} \sin \theta^{\icd}_{k} \Big] ^{-j \kappa (y\sin \theta^{\icd}_{k} - z\cos \theta^{\icd}_{k})},
\end{align}
where $\eta = {\omega} / {\kappa}$. Following the analysis in \cite[Section 11.3.2]{BalanisWiley2012} and \cite{DovelosICC2021}, the corresponding power density of the scattered E-field at $\rx_l$ is obtained as
\begin{align}
S^{\rfl}_{\tx_k, \IRS_n, \rx_l} &= 
    \frac{p_k \gain_{\tx_k}}{4 \pi}
    \bigg( \frac{\frac{A}{\lambda}}{d_{\tx_k, \IRS_{n,m}} d_{\IRS_{n,m}, \rx_l}} \bigg)^2\nonumber\\&\quad\times
    F(\psi_{\tx_k, \IRS_{n,m}},\phi_{ \IRS_{n,m}, \rx_l },\psi_{ \IRS_{n,m}, \rx_l }).
\end{align}

Hence, considering the molecular absorption loss in THz communication and the receive aperture $\frac{\gain_{\rx_l} \lambda^2}{4\pi}$, the received signal at $\rx_l$ can be expressed as
\begin{align} \label{eq_rx_signal_v1}
y  = \sum_{m=1}^{M_n} 
    &\sqrt{ \frac{\gain_{\rx_l} \lambda^2}{4\pi} S^{\rfl}_{\tx_k, \IRS_n, \rx_l} } \kappa_{n, m} 
    \sqrt{e^{-\kappa_{\sf abs}(f_c)(d_{\tx_k,\IRS_{n,m}}+d_{\IRS_{n,m},\rx_l})}} 
     \nonumber\\&\times 
    e^{ - j \frac{2 \pi}{\lambda} (d_{\tx_k,\IRS_{n,m}}+d_{\IRS_{n,m},\rx_l})} 
    e^{ j\theta_{n,m} } x_k + n_{\rx_l},
 \end{align} 
where 
    $F(x,y,z)= \cos^2 x (\cos^2 y \cos^2 z +\sin^2 y)$, $\kappa_{n,m} \in (0,1]$ and $\theta_{n,m} \in [0, 2\pi)$ represent the amplitude and the PS of $\IRS_{n,m}$, respectively. Moreover,
    $\kappa_{\sf abs}(f)$ $[\textnormal{m}^{-1}]$ is the molecular absorption coefficient at the frequency of interest $f$ [Hz].
For $f \in [275, 400]$ GHz, $\kappa_{\sf abs}(f)$ can be determined as \cite{TarboushTVT2021}
\begin{align}
\kappa_{\sf abs}(f)
    =   y_1(f, \mu_{\textnormal{H}_2 \textnormal{O}})
        +   y_2(f, \mu_{\textnormal{H}_2 \textnormal{O}})
        +   g(f),
\end{align}
where $y_1(f, \mu_{\textnormal{H}_2 \textnormal{O}})$, and $y_2(f, \mu_{\textnormal{H}_2 \textnormal{O}})$, $g(f)$ are given by \cite[Eq. (31), Eq. (32), Eq. (33)]{TarboushTVT2021}, respectively, $\mu_{\textnormal{H}_2 \textnormal{O}} = \frac{\phi}{100} \frac{p^\ast_w(T, p)}{p}$ is the volume mixing ratio of water vapor, $\phi \in [0, 100]$ is the relative humidity, and
    $\frac{\phi}{100} p^\ast_w(T, p)$ is the partial pressure of water vapor at temperature $T$ [K] and pressure $p$ [hPa]. 
Specifically, $p^\ast_w(T, p)$ is given by
\begin{align}
p^\ast_w(T, p) = 6.1121 \bigg(1.0007+ \frac{3.46}{10^6} p\bigg)
    e^{ \frac{17.502 (T-273.15)}{(T-32.18)} }.
\end{align}

Henceforth, we consider the standard atmospheric condition, where $T = 296$ K, $p = 1013.25$ hPa, and $\phi = 50$, denoting $50\%$ humidity. Furthermore, the received signal corresponding to~\cite{OzcanTVT2021}, can be expressed as
\begin{align} \label{eq_rx_signal_v2}
y  = \sum_{m=1}^{M_n} &\frac{\sqrt{p_{k}A^2 \gain_{\tx_k} \gain_{\IRS_{n,m}}(-\vec{r}_{\tx_k, \IRS{_n,m}})}}{4 \pi d_{\tx_k,\IRS_{n,m}} } \sqrt{e^{-\kappa_{\sf abs}(f_c)d_{\tx_k,\IRS_{n,m}}}}\nonumber\\&\quad\times 
\frac{\sqrt{A^2 \gain_{\rx_l} \gain_{\IRS_{n,m}}(\vec{r}_{\IRS_{n,m}, \rx_l})}}{4 \pi  d_{\IRS_{n,m},\rx_l}} \sqrt{e^{-\kappa_{\sf abs}(f_c)d_{\IRS_{n,m},\rx_l}}} 
\nonumber \\
& \quad \times 
e^{ - j \frac{2 \pi}{\lambda} (d_{\tx_k,\IRS_{n,m}}+d_{\IRS_{n,m},\rx_l})} 
e^{ j\theta_{n,m} } x_k + n_{\rx_l}.
\end{align} 
Comparing~\eqref{eq_rx_signal_v1} and~\eqref{eq_rx_signal_v2}, we can formulate the gain of the $m^{th}$ element of $\IRS_n$ as \eqref{gain}. This completes the proof of Lemma \ref{lemma_gain}.
\section{{Proof of Lemma~\ref{lemma_maxmin}}}\label{ap_lemma2}
In the first phase of our proposed algorithm, we find the Tx to IRS matching to maximize the pseudo quantity from Tx to IRS, which can be written as $ [\vec{\overline{\Omega}}]_{(\tx_k^\star,\IRS_n^\star, :)} = \arg \max_{\tx_k,\IRS_n, :}  \{ \Lambda_{\tx_k,\IRS_n, :} \},$ where $\tx_k^\star$ and $\tx_n^\star$ are the best possible allocation to maximize the pseudo quantity. The allocation achieved in the first phase is utilized to find the allocation matrix between the IRS and the receiver in the second phase, as $[\vec{\overline{\Omega}}]_{\tx_k,\IRS_n,\rx_l}^\star = \arg \max_{\tx_k^\star,\IRS_n^\star,\rx_l} \left\{ \Lambda_{\tx_k^\star,\IRS_n^\star,\rx_l} \right\}$.
The sum rate achieved by the proposed method is upper bounded by the sum rate obtained with the ES method, which can be expressed as
\begin{align}
& \max\limits_{\tx_k,\IRS_n,\rx_l} \Big \{\sum_{\tx_k,\IRS_n,\rx_l}  R_{\tx_k,\IRS_n,\rx_l} \Big\}\geq \max\limits_{\tx_k^\star,\IRS_n^\star,\rx_l} \Big \{\sum_{\tx_k^\star,\IRS_n^\star,\rx_l} \Lambda_{\tx_k^\star,\IRS_n^\star,\rx_l} \Big\},
\end{align}
which is equivalent to
\begin{align}
   &\max\limits_{\tx_k,\IRS_n,\rx_l}  \big\{\sum_{\tx_k,\IRS_n,\rx_l}   \log_{2} (1 + \SINR_{\tx_k,\IRS_n,\rx_l}) \big\}\nonumber\\&\quad\geq \max\limits_{\tx_k^\star, \IRS_n^\star, \rx_l} \big\{\sum_{\tx_k^\star,\IRS_n^\star,\rx_l}  \log_{2} (1 + \Xi_{\tx_k^\star,\IRS_n^\star,\rx_l}) \big\}.
\end{align}
Since the $\log$ function is a monotonically increasing function, we have
\begin{align}
    &\max\limits_{\tx_k,\IRS_n,\rx_l} \Big\{\sum_{\tx_k,\IRS_n,\rx_l} \SINR_{\tx_k,\IRS_n,\rx_l} \Big\} \geq \max\limits_{\tx_k^\star,\IRS_n^\star,\rx_l} \Big\{\sum_{\tx_k^\star,\IRS_n^\star,\rx_l} \Xi_{\tx_k^\star,\IRS_n^\star,\rx_l} \Big\}.
\end{align}
Using ideal PS, the maximum SINR can be obtained from~\eqref{eq_sinr_e2e2}. When $\tx_k$, $\IRS_n$, $\rx_l$ and $\tx_k^\star$, $\IRS_n^\star$, $\rx_l$ are similar, the proposed sub-optimal solution gives the same rate as the ES and
\begin{align}\label{equi}
   &\max\limits_{\tx_k,\IRS_n,\rx_l} \Big\{\sum_{\tx_k,\IRS_n,\rx_l} \SINR_{\tx_k,\IRS_n,\rx_l} \Big\}= \max\limits_{\tx_k,\IRS_n,\rx_l} \Big\{\sum_{\tx_k,\IRS_n,\rx_l} \Xi_{\tx_k,\IRS_n,\rx_l} \Big\}.
\end{align}
We consider the right side of~\eqref{equi}, for the same allocation matrix with ideal PS using~\eqref{eq_sinr_sr} and~\eqref{eq_sinr_ru}, $\{\Xi_{\tx_k,\IRS_n,\rx_l}\}$ is given in~\eqref{eq_sinr_proof}, at the top of next page, which is equivalent to the SINR as in~\eqref{eq_sinr_e2e2}.
\begin{figure*}[t]
 	\begin{align} \label{eq_sinr_proof}
		 \Xi_{\tx_k,\IRS_n,\rx_l} 
		\!\!\!=\frac{\frac{p_k\lambda^2}{16\pi^2}\Big|\sum\limits_{m=1}^M \frac{ \sqrt{ \gain_{\tx_k} \gain_{\rx_l} \gain_{\IRS_{n,m}}(-\vec{r}_{k,n,m})\gain_{\IRS_{n,m}}(\vec{r}_{n,m,l}) e^{-\kappa_{abs}(f)(d_{\tx_k,\IRS_{n,m}}+d_{\IRS_{n,m},\rx_l})}}}{ d_{\tx_k,\IRS_{n,m}} d_{\IRS_{n,m},\rx_l}}
			\Big|^2}{\sum\limits_{\substack{j=1\\ j\neq k}}^K \sum\limits_{\substack{i=1}}^N p_j\big|(\widehat{\vec{h}}_{\tx_j,\IRS_i})^T\vec{\Theta}_i\widehat{\vec{g}}_{\IRS_i,\rx_l}\big|^2\!\!+\!\sum\limits_{\substack{j=1}}^K \sum\limits_{\substack{i=1}}^N p_j\big(\sigma_{\widetilde{g}_{\IRS_i,\rx_l}}^2\big|\vec{\widehat{h}}^T_{\tx_j,\IRS_i}\vec{\widehat{h}}_{\tx_j,\IRS_i}\big|+\sigma_{\widetilde{h}_{\tx_j,\IRS_i}}^2\big|\vec{\widehat{g}}^T_{\IRS_i,\rx_l}\vec{\widehat{g}}_{\IRS_i,\rx_l}\big|+\sigma_{\widetilde{h}_{\tx_j,\IRS_i}}^2\sigma_{\widetilde{g}_{\IRS_i,\rx_l}}^2\big)\!+\!\sigma_l^2}.
	\end{align}
	\hrule 
\end{figure*}
 This completes the proof of Lemma \ref{lemma_maxmin}.
\section{Proof of Lemma~\ref{lemma3}} \label{ap_lemma3}
To prove Lemma~\ref{lemma3}, we use induction on the subsequent iteration $j$, where $j\geq i$. The inductive reasoning is as follows:

Initial step ($j=i$): At the $i^{th}$ iteration, \textit{\textbf{proposer}} $P$ is proposing to \textit{\textbf{responder}} $R$ and two cases are possible according to our algorithm: $(i)$ Either $R$ is free, in which case $R$ accepts the proposal from $P$, or $(ii)$ $R$ is paired with another \textit{\textbf{proposer}} $P^\star$, and $R$ accepts $P$ only if $R$ prefers $P$ over $P^\star$. Thus, in either cases, on the $i^{th}$ iteration, $R$ pairs with the \textit{\textbf{proposer}} which is at least as good as $P$.

\textit{Hypothesis:} Assuming the claim in initial step that $R$ pairs with the \textit{\textbf{proposer}} which is at least as good as $P$ is true for $j\geq i$.

\textit{Induction step:} We prove the claim for the $(j+1)^{th}$ iteration. According to the above hypothesis, on $j^{th}$ iteration, $R$ is paired with  $P^\star$, which it prefers at least as much as $P$. At the $(j+1)^{th}$ iteration, two cases are possible: $(i)$ Either there are no new \textit{\textbf{proposers}} for $R$, and $R$ continue with $P^\star$, or $(ii)$ a new \textit{\textbf{proposer}} $P^{\star\star}$ proposes to $R$, and $R$ accept the new \textit{\textbf{proposer}} only if $R$ prefers $P^{\star\star}$ over $P^\star$. In both cases, $R$  pairs with the \textit{\textbf{proposer}} it prefers at least as much as $P$.
\section{{Proof of Theorem~\ref{theorem2}}}\label{ap_theorem2}
Let $\mathcal{P}$ and $\mathcal{R}$ denote the set of all proposers and responders, respectively. We consider a bipartite graph $G$ whose vertex partitions correspond to $\mathcal{P}$ and $\mathcal{R}$, and there is an edge between a proposer $p \in \mathcal{P}$ and a responder $r\in \mathcal{R}$. 
 We prove that there is a stable matching obtained by the proposed algorithm using contradiction as follows. Suppose that some \textit{\textbf{proposer}} $p \in \mathcal{P}$ is rejected by the best valid \textit{\textbf{responder}} $r=best(p)$ ($r \in \mathcal{R}$) in the proposed algorithm. Furthermore, $r$ rejects $p$ in favor of $p^\star \in \mathcal{P}$, which $r$ likes more $p^\star$ than $p$. Thus, in $r's$ priority, we have $p^\star \succ^{{r}} p$. Let us indicate this point as $\textit{\textbf{I}}$ (rejection point), and return to it later to drive a contradiction. According to the hypothesis that $r=best(p)$, there exists an edge between vertex $p$ and $r$ in $G$ which make a stable matching $\Omega'$. Moreover, in $\Omega'$, there is an edge between $p^\star$ and $r^\star \in R$ in a bipartite graph $G$. Thus, $\Omega'$ contains stable matches ($p, r$) and ($p^{\star}, r^\star$). We now consider what the implementation of the proposed algorithms decides about the priorities of $p^\star$ between $r$ and $r^\star$. Since $\textit{\textbf{I}}$ was the first event in the algorithm where any \textit{\textbf{proposer}} was rejected by its best valid partner, at this instant the following must be true: $(i)$ $p^\star$ has not been rejected by its $best(p^\star)$, and  $(ii)$ $p^\star$ has been rejected by every \textit{\textbf{responder}} in the list that comes before $r$, since $p^\star$ is paired with $r$ at instant \textit{\textbf{I}} as proved in lemma 4.
The above two facts jointly imply that $r \succ^{{p^\star}} r^\star$ in the priority list of $p^\star$. First, $best(p^\star)$ must come after $r$ because prior hasn't rejected $p^\star$ yet, as proof in lemma 4, so $r > best(p^\star)$. Secondly, $r^\star$ is always a valid match with $p^\star$ because $\Omega'$ is a stable matching. This means ($p^\star, r$) is an unstable match for $\Omega'$; however, in $\Omega'$ both $p^\star$ and $r$ prefer each other to their assigned pairs. Thus, our initial assumption that $p$ was rejected by $best(p)$ is false, which proves that the algorithm returns the stable matching.

Now, we prove that the stability does not change when inverting the proposing order (i.e., previous \textit{\textbf{proposer}} as \textit{\textbf{responder}} and \textit{\textbf{responder}} as \textit{\textbf{proposer}}). The cascaded channel from the $\tx$-to-$\IRS$-to-$\rx$ is composed of three parts, namely the channel from the $\tx$-to-$\IRS$, the PS matrix of the IRS, and the channel from the $\IRS$-to-$\rx$. In our proposed work, we consider the channel from the $\tx$-to-$\IRS$ in the first phase, while in the second phase, the channel from the $\IRS$-to-$\rx$ is considered. Moreover, channel reciprocity holds for the $\tx\to\IRS$ and $\IRS\to\rx$ channels~\cite{tang2021channel}. Thus, inverting the proposing side doesn't affect the stability of the proposed algorithms.
\bibliographystyle{IEEEtran}
\balance
\bibliography{main}

\end{document}